\documentclass[preprint,apl]{revtex4-1}
\usepackage{amsmath, amssymb}
\usepackage{makeidx}         
\usepackage{graphicx,epsf,epsfig}        
\usepackage{graphicx}   
\usepackage[bottom]{footmisc}
\usepackage{epstopdf}

\usepackage{color}

\makeindex             

\begin{document}
\title{Graphene Nanoribbons: from chemistry to circuits}
\author{F. Tseng } \author{D. Unluer} \author {M. R. Stan} \author{A. W. Ghosh}
\affiliation{Charles L Brown School of Electrical and Computer Engineering, University of
Virginia, Charlottesville, VA22901
\texttt{ft8e@virginia.edu}}

\begin{abstract}
The Y-chart is a powerful tool for understanding the relationship between various views (behavioral, structural, physical) of a system, at different levels of abstraction, from high-level, architecture and circuits, to low-level, devices and materials. We thus use the Y-chart adapted for graphene to guide the chapter
and explore the relationship among the various views and levels of abstraction.
We start with the innermost level, namely, the structural and chemical view.
The edge chemistry of patterned graphene nanoribbons (GNR) lies intermediate between graphene and benzene, and the corresponding strain lifts the degeneracy
that otherwise promotes metallicity in bulk graphene. At the same time, roughness at the edges washes out chiral signatures, making the nanoribbon width the principal arbiter of metallicity. The width-dependent conductivity allows the design of a monolithically patterned wide-narrow-wide all graphene interconnect-channel heterostructure. In a three-terminal incarnation, this geometry exhibits superior
electrostatics, a correspondingly benign short-channel effect and a reduction in the contact Schottky
barrier through covalent bonding. However, the small bandgaps make the devices transparent to band-to-band tunneling. Increasing the gap with width confinement (or other ways to break the sublattice
symmetry) is projected to reduce the mobility even for very pure samples, through a fundamental {\it{asymptotic}}
constraint on the bandstructure. An analogous trade-off, ultimately between error rate (reliability) and delay (switching speed) can be projected to persist for all graphitic derivatives. Proceeding thus to a higher level, a compact
model is presented to capture the complex nanoribbon circuits, culminating in inverter characteristics, design metrics and layout diagrams. 
\end{abstract}
\maketitle
\begin{figure}[h]\begin{tiny}
\end{tiny}
\centering
\includegraphics*[width=0.9\textwidth]{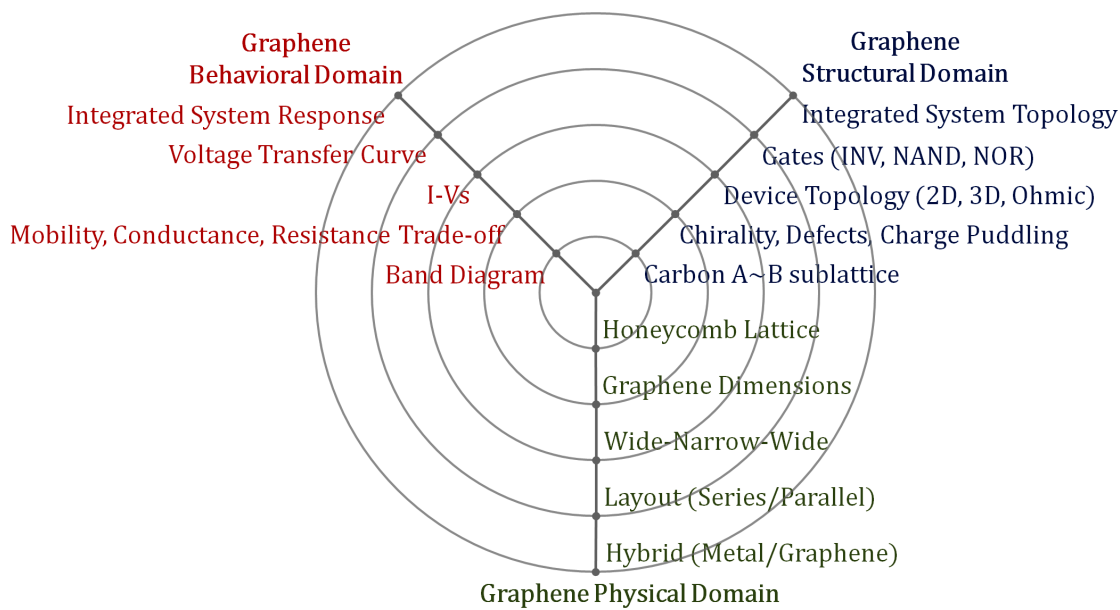}
\caption[]{Gajski-Kuhn Y-chart adapted for graphene}
\label{fig:chart}
\end{figure}

The Gajski-Kuhn Y-chart (Fig.~\ref{fig:chart}) is a model which captures in a snapshot view, the essential considerations in designing semiconductor devices ~\cite{dos01}. The three domains of the Gajski-Kuhn Y-chart are on radial axes.  Each of the domains can be divided into levels of abstraction, using concentric rings.  At the top level (outer ring), we consider the architecture of the chip; at the lower levels (inner rings), we successively refine the design into finer detailed implementation:

\begin{itemize}
\renewcommand{\labelitemi}{$\bullet$}
\item Creating a structural description from a behavioral one is achieved through the
processes of high-level synthesis or logical synthesis.
\item Creating a physical description from a structural one is achieved through layout synthesis.
\end{itemize}

Building up such a multilevel view of graphene, a material of undoubted  interest,
is an ongoing process. The aim of this chapter is to touch upon just a few of the
concentric circles to create such a viewing platform.

\section*{18.1. The innermost circle -- the atomistic view}
Although there has been a lot of effort at this end from physicists and chemists,
our aim here is more object-oriented -- we wish to touch upon the GNRs, specifically,
the edge chemistry that leads to the observed lack of chirality and metallicity,
then progressively outward along the Y-chart towards their two and then three
terminal properties, and finally to an overall compact model describing its circuit
level potential.

\subsection*{18.1.1 Flatland: A romance in two dimensions}

Among elements on the periodic table, carbon is unique in that its sp$^2$ planar and sp$^3$ tetragonal bond energies are comparable, making it geometrically vary all the way from 3-d diamond to 0-d
buckyballs ~\cite{rev01}. Of this entire set, a candidate that combines impressive
material properties (e.g. ultra-low effective mass), structural versatility, and amenability to
industry-standard planar fabrication techniques, is undoubtedly graphene and its multiple derivatives, including
carbon nanotubes (CNTs), bilayer graphene (BLG), epitaxial graphene (epi-G), strained graphene (sG) and graphene
nanoribbons (GNR).

The impressive electronic properties of CNTs and GNRs stem from their parent graphitic bandstructure ~\cite{che01}. Without delving into the mathematics, it is important to
rehash some of the salient features relating the bandstructure to the underlying chemistry.
The hybridization of one $s$ and two carbon $p_{x,y}$-orbitals creates a planar honeycomb structure in a single graphene sheet, loosely resembling self-assembled benzene
molecules minus the hydrogen atoms. The crystal structure can be described as a triangular network with a two-atom dimer basis, whose $\pi$ electrons hybridize to create bonding-antibonding pairs that delocalize over the entire crystal to generate conduction and valence bands. However, since the two basis atoms and the orbitals involved are identical, we get a zero-band gap metal with a dispersion resembling photons, albeit with a much lower speed. The resulting low energy linear dispersion corresponds to a constant slope and thus a constant velocity $v = \partial E/\hbar\partial k$.

The unique bandstructure of graphene contributes to its amazing electronic properties ~\cite{rev01}.
Because the Fermi velocity is energy-independent, the cyclotron effective mass of graphene electrons, $m^* = \hbar k_F/v_F$ is vanishingly small at low energy ($k_F \rightarrow 0$, $v_F$ being constant at roughly $10^8$ cm/s). Furthermore, the two bands are derived out of symmetric and antisymmetric (bonding and antibonding) combinations of the two identical dimer atoms, creating a two component pseudo-spinor out of the two Bloch wavevectors, with their ratio being just a phase factor $e^{i\theta}$, where $\tan\theta = k_y/k_x$ relates electron quasi-momentum components in the graphene x-y plane. The reversal of phase between the forward and backward velocity vectors suppresses 1D acoustic phonon back-scattering, allowing only  Umklapp processes in confined graphitic structures such as CNTs and GNRs.
The combination of low mass $m^*$ and high mean free path $\lambda$ ultimately
leads to very high mobilities $\mu = q\lambda/m^*v_F$, with a record room-temperature
value at 230,000 $cm^2/Vs$ ~\cite{bol01} for suspended graphene sheets.

In the following sections, we discuss two bandstructure related issues that arise when we attempt to pattern or modify graphene to generate gapped or confined planar structures -
\begin{itemize}
\renewcommand{\labelitemi}{$\bullet$}
\item the absence of metallicity and chirality dependent bandgaps
in multiple experiments (sections 18.1.2-18.1.4), and
\item the increase in effective mass as a gap is progressively opened, arising from fundamental
asymptotic constraints (section 18.1.5) that are expected to persist even for very pure samples.
\end{itemize}

\subsection*{18.1.2 Whither metallicity?}

The bandstructure of CNTs is by now, a textbook homework problem. Since
the conduction and valence bands of its parent graphene structure touch
precisely at the vertices of its hexagonal Brillouin zone, the Fermi
wavelength of undoped graphene corresponds to a {\it{unique electron wavelength}}
$\lambda_F = 3\sqrt{3}R/2$, where $R$ is the C-C bond distance. The
imposition of periodic boundary conditions along the circumference of a
CNT filters out many allowed electron wavelengths and allows only modes that
have integer ratios of $\pi D/\lambda$ ($D$ being the tube diameter), so that a particular
chirality (i.e., a wrapping topology) may or may not support $\lambda_F$
needed to sustain metallicity. Accordingly, one can derive selection rules
on paper, or using simple 1-orbital orthogonal nearest neighbor tight binding,
that matches experiments quite well. In any random array of CNTs, roughly
a third are expected to be metallic, and impressive progress has been achieved in sorting them out from their semiconducting counterparts. Even in extremely narrow CNTs with
strong curvature-induced out-of-plane hybridizations, the anomalous bandgaps
are well captured by non-orthogonal tight-binding formulations such as Extended
H\"uckel Theory ~\cite{kie01}. The bandgaps of CNTs bear relatively few surprises, at the
end of the day.

\begin{figure}[ht]
\includegraphics*[width=1.0\textwidth]{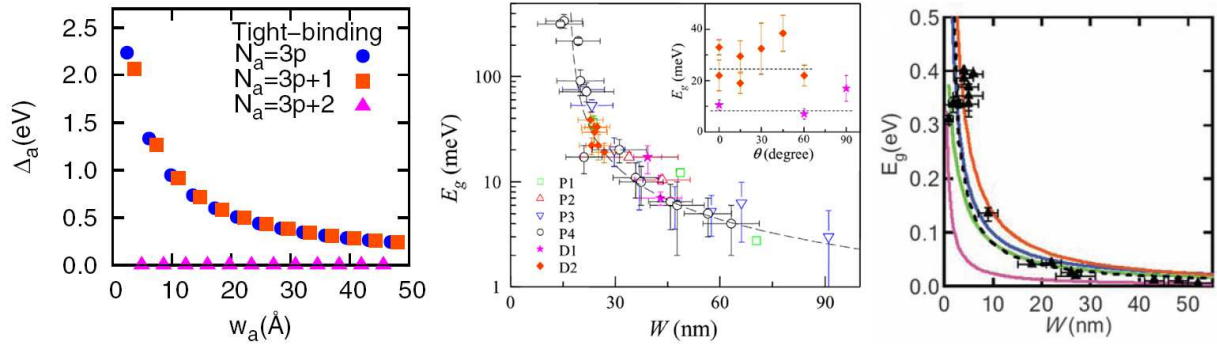}
\caption{(Left) Tight-binding 1-orbital calculations show three chiral curves, one of which is metallic ~\cite{son01}. In contrast, data from (center) Philip Kim's group ~\cite{han01} and (right) Hongjie Dai's group ~\cite{dai02} show a single chirality free curve with no metallic signatures.}
\label{nochirality}
\end{figure}

Life is more complicated when dealing with GNRs. Indeed, it seems reasonable to expect
a chirality dependence to arise for GNRs, simply replacing the periodic circumferential boundary conditions across CNTs with hard-wall boundary conditions across the GNR width.
A few details may change, for instance, we now fit half-wavelengths rather than whole wave-lengths, and the edges are not completely opaque to electrons tunneling outwards so that the boundary conditions are more `diffuse'. The quantization condition will roughly correspond now to integer ratios for $(W+R\sqrt{3})/\lambda$ (accounting for two unit cells outside for the wavefunction to vanish).
However, one-orbital tight binding would still predict three chiral classes in GNRs, one of which is strictly metallic as in CNTs.
Experimentally however, no chirality dependence is observed for GNRs, nor are any GNRs observed to be strictly metallic at low temperature (Fig.~\ref{nochirality}). Regardless of wrapping vector,
GNRs wider than 10 nm are quasi-metallic while narrow ones semiconducting. We thus have a startling disconnect between simple theories and experiments ~\cite{dai02}.

\subsection*{18.1.3. Edge chemistry -- benzene or graphene?}

\begin{figure}[h!]
\center
\includegraphics*[width=1\textwidth]{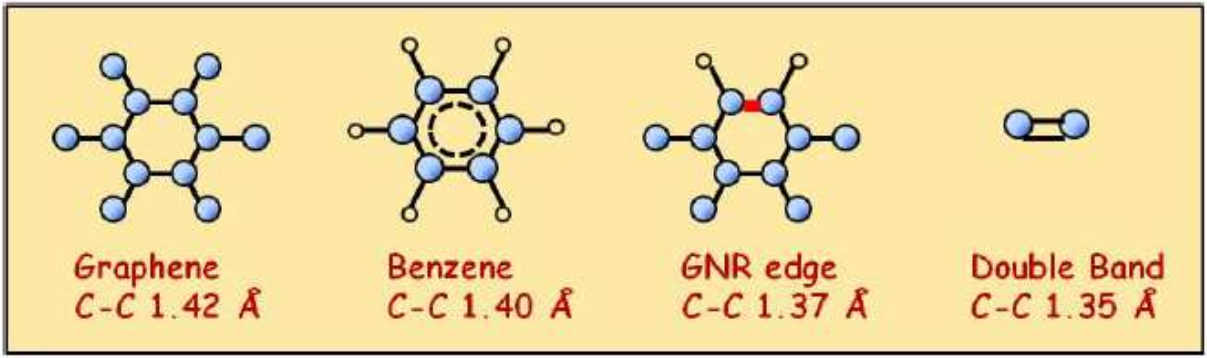}
\caption{C-C bond length comparisons show that relaxed armchair GNR edges lie between benzene and double bonds, enjoying only a partial resonant hybridization.} \label{graben}
\end{figure}

The disconnect between naive expectations and observations arises from the
boundaries, which ultimately impact the quantization rules behind patterned GNRs.
Specifically, we argue that edge strain and roughness are the main factors
behind the disconnect. Doing justice to such effects will need a proper bandstructure
that can capture atomistic chemistry and distortion. While
Density Functional Theory (DFT) within the LDA-GGA approximation captures these
effects well, we used DFT primarily for structure evaluation, and resorted to
Extended H\"uckel theory (EHT) fitted to
bulk graphene to explore the low-energy bandstructures.
We relaxed the hydrogenated edges of armchair GNRs using LDA-GGA and found
a bonding environment distinct from bulk graphene. While the inner carbon atoms have a bond-length of 1.42$\AA$, the edges tend to dimerize and see a 3.5$\%$ ~\cite{son01} ~\cite{gun01} strain associated with a reduction in bond length to 1.37$\AA$. Thereafter, we employ non-orthogonal basis sets in EHT to capture the effect of the edge chemistry on the low-energy electronic structure.

There is an appealingly simple explanation for the observed bonding chemistry. Since the edge atoms are connected to hydrogen on one side and carbon on the other, the difference in electronegativity tends to move the edge C-C atoms closer to a benzene structure (Fig.~\ref{graben}). However, the unequal bonding environment at the edge disallows any resonant hybridization that evens out the double-single bond distribution in aromatic rings, so that the edge rings in GNRs break into `domains' with nearly intact double bonds at the edges and slightly expanded single bonds towards the bulk end. Since benzene is semiconducting, the 3.5$\%$ strain at the dimerized edges increases intradimer overlap but reduces interdimer overlap, effectively opening a bandgap by 5$\%$. The obvious consequence is that {\it{all armchair GNRs become strictly semiconducting}}, in sharp contrast to their CNT counterparts. In contrast, the edges of zigzag GNRs have a lateral symmetry that makes them resistant to dimerization. In fact, the inward motion of the C-C edges away from the hydrogens shrinks both bond lengths equally, making zigzag edges more conducting.

\begin{figure}[t]
\includegraphics[scale=0.1052]{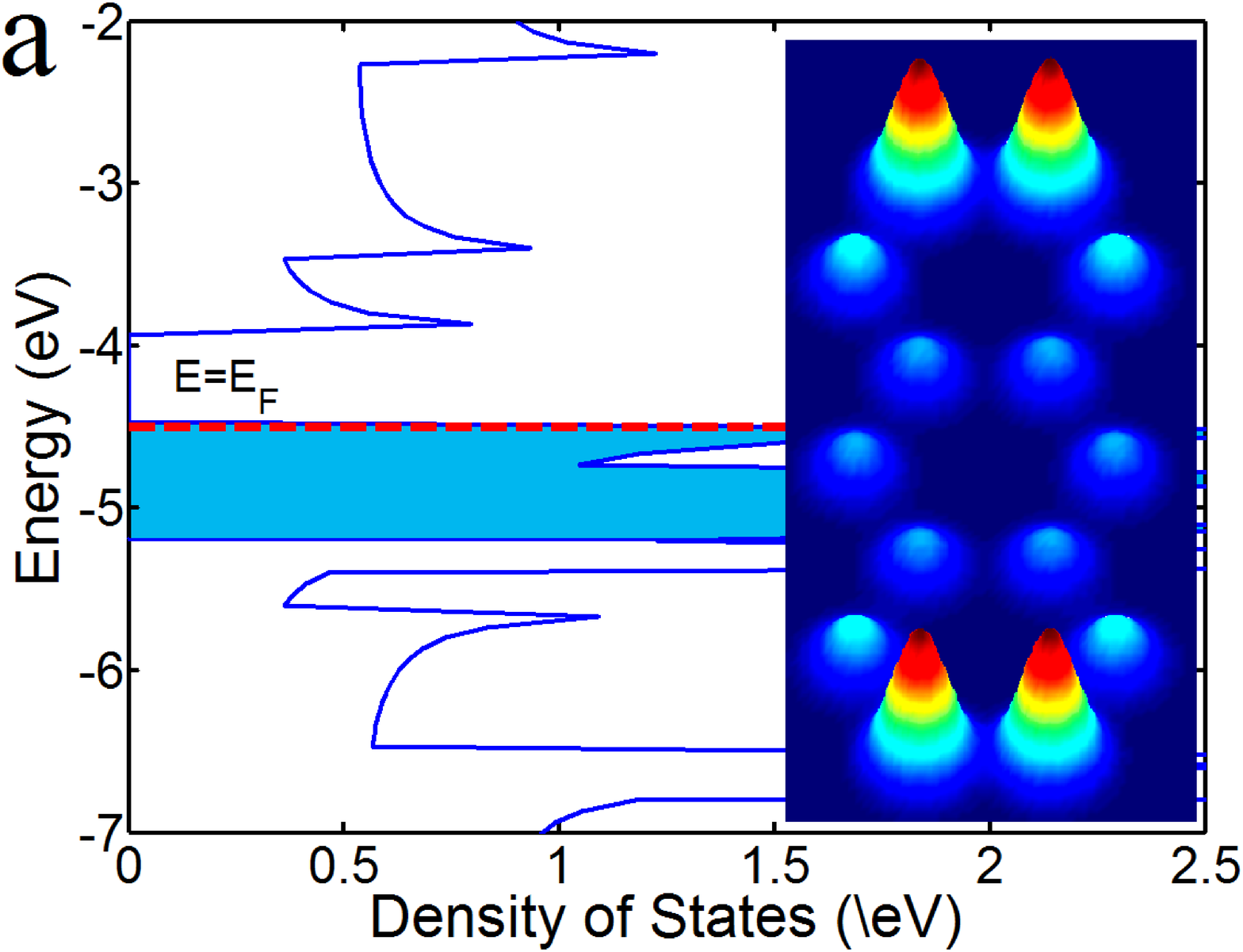}\label{a}
\hskip 2cm\includegraphics[scale=0.1052]{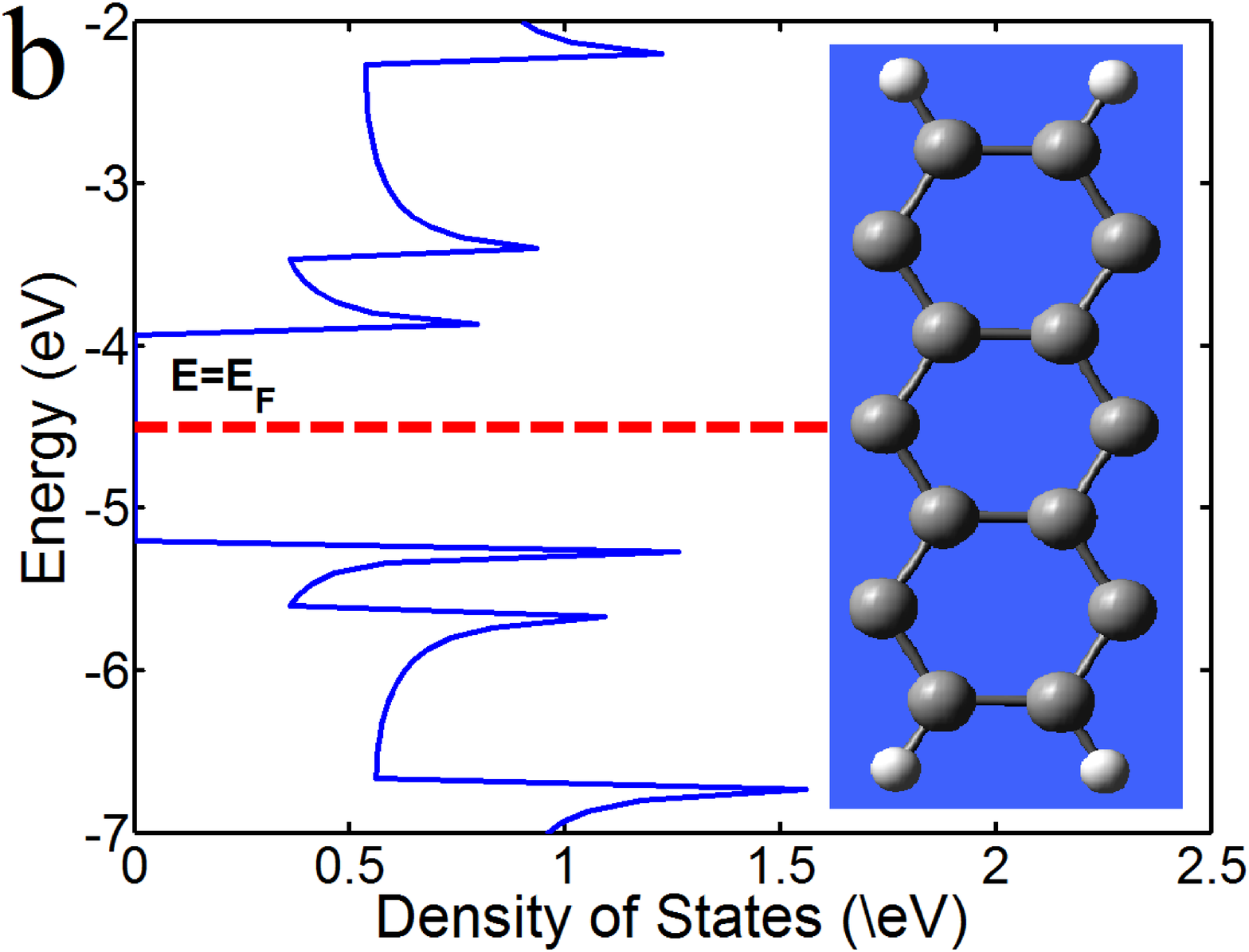}\label{b}
\vskip 1cm
\includegraphics[scale=.1051]{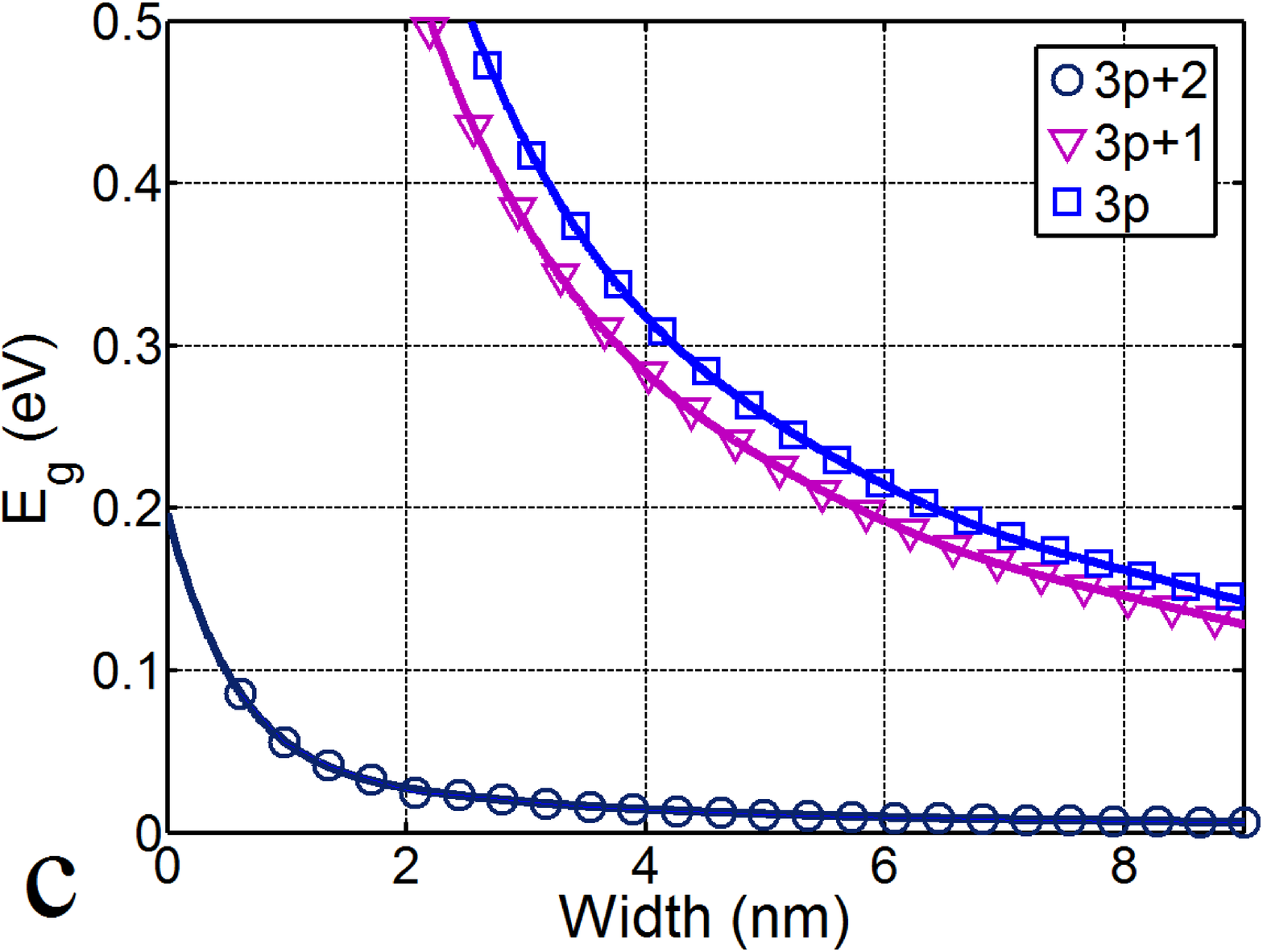}\label{c}
\hskip 2cm\includegraphics[scale=.1050]{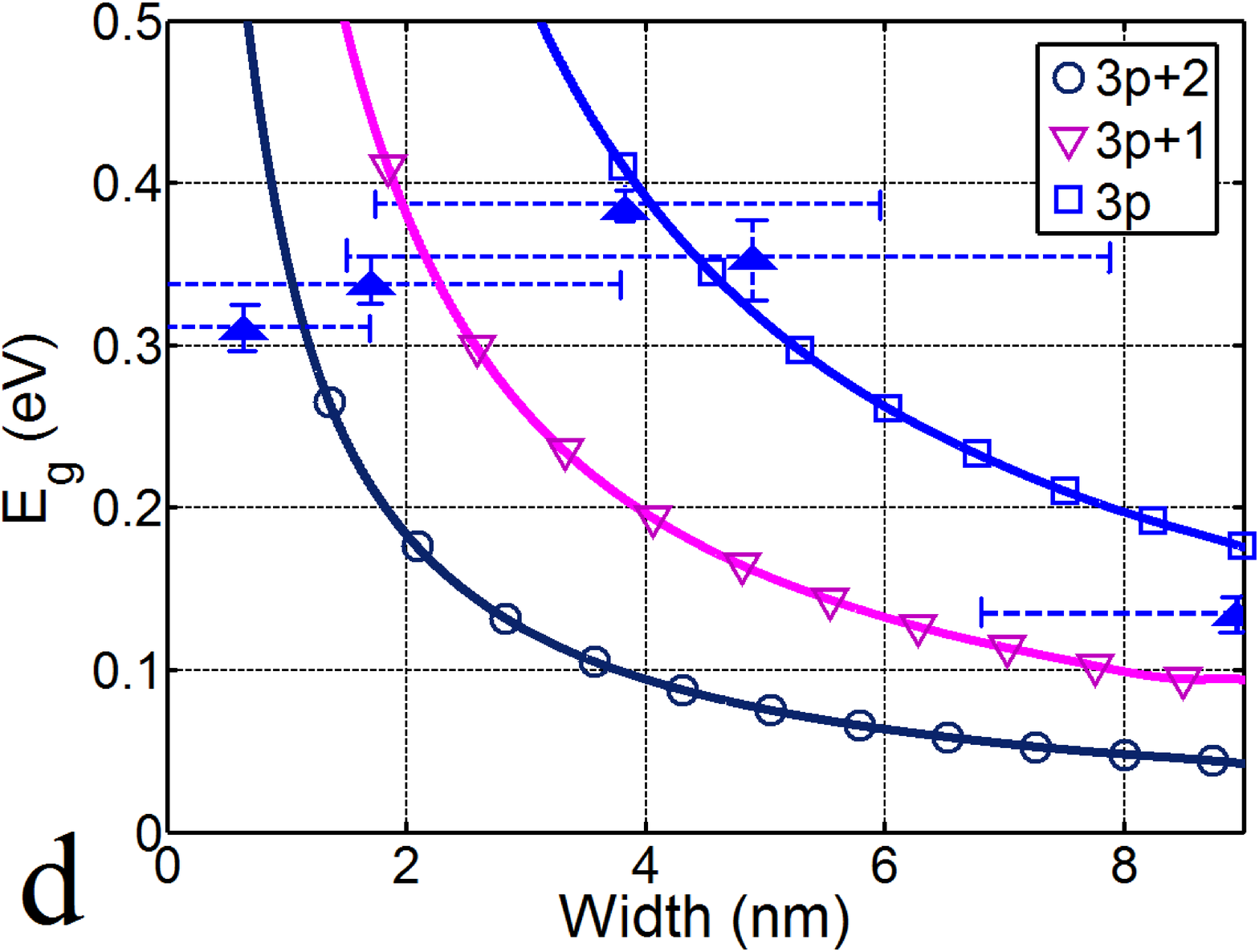}\label{d}
\caption[]{(a) Open carbon bonds at the edges introduce edge-states (shaded) in the DOS.  Spatial resolution of those eigenstates around the Fermi energy confirms the electron wavefunction localized at the armchair edges.
(b) When open carbon bonds are hydrogen terminated, those edge-states are removed.
(c) Applying EHT to GNR dispersion relation across a range of sub-10nm armchair edge widths finds an oscillating bandgap.
(d) Relaxation of edge bonds that are hydrogen terminated widens the energy bandgap for \textit{$3p$} and reduces the gap for \textit{$3p+1$} GNRs. $E_{g}$ vs width results are within range of experimental data points \cite{dai02} and also in agreement with DFT predictions.}
\vspace{-0.2 in}
\label{ehtstrain}
\end{figure}

To calculate the impact of the relaxed bonds on the electronic structure, we calculated the density of states of a uniformly wide armchair GNR with edge relaxation using Extended H\"uckel Theory. While EHT has been used widely to study molecular properties, it has also been extended to describe bulk semiconducting bandstructures using localized Wannier like non-orthogonal basis sets that still retain their individual orbital properties. Through extensive tests on both graphene and silicon, we have found that EHT accurately captures both bulk bandstructures as well as surface and edge distorted bandstructures ~\cite{tse01}. The result of our simulation is shown in Fig.~\ref{ehtstrain}. The role of edge passivation is shown at the top, where we can see the explicit removal of edge induced midgap states by hydrogenation. The bottom panels show the role of edge strain. In contrast to $p_z$ based nearest neighbor one orbital tight-binding theory, a small bandgap opens ~\cite{kie01}. While CNTs have precise periodic boundary conditions along their circumference, the edge atoms do not provide an exact hard wall boundary condition, as the electrons tend to tunnel out into the surrounding region. In the presence of edge relaxation, the gap increases because of the aforementioned dimerization, removing any semblance of metallicity from the bandgap vs width plots.

\begin{figure}[ht]
\centering
\includegraphics*[width=.9\textwidth]{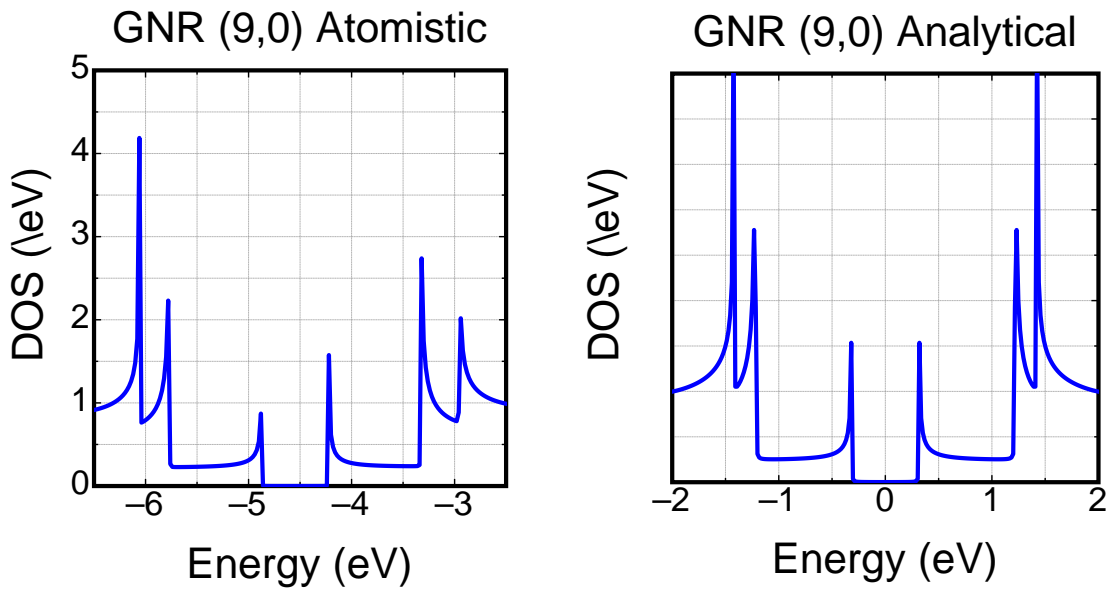}
\caption{Atomistic EHT plot (left) of the DOS of a (9,0) relaxed armchair GNR, compared with a next-nearest neighbor tight-binding fit (right) from C. T. White at NRL. From the $p_z$ sector of the EHT Hamiltonian, we estimate the nearest neighbor coupling, third nearest neighbor coupling and the shift in coupling at the $3.5\%$ strained edges as $t_1 = -3.31$ eV, $t_3 = -0.106$ eV and $\Delta t_1 = -0.38$ eV. These extracted parameters compare quite favorably with C. T. White's tight binding parameters, $t_1 = -3.2$ eV, $t_3 = -0.3$ eV and $\Delta t_1 = -0.2$ eV. ~\cite{gun01}} \label{ehtwhite}
\end{figure}

While EHT explains the removal of metallicity, compact models prefer a suitably
calibrated orthogonal tight-binding model, with the edge chemistry
through beyond nearest neighbor interactions. It is not clear if this reproduces the
Bloch wavefunctions, but they do seem to capture the overall density of states.
Fig.~\ref{ehtwhite} shows a comparison between a (9,0) EHT DOS and a (9,0) tight-binding Density of States (DOS) (parametrized independently by CT White) ~\cite{are01}. We will use this formulation for our simpler compact models described later.

\subsection*{18.1.4. Whither chirality?}

While we can explain away the lack of metallicity through the preponderance of
edge strain, why do we not see the three chiral curves in a plot of experimentally measured bandgaps versus ribbon widths? The primary reason, we believe, is roughness at the edges, which tends to wash away such chiral signatures.
Currently, line edge roughness along a GNR edge is an unavoidable consequence of lithographic and chemical fabrication techniques ~\cite{bil01} ~\cite{wan01}. Unzipping CNTs via ion bombardment produces the smoothest GNR edges to date ~\cite{kos01}. However edge fluctuations even on the scale of a single atom can degrade transmission probability of modes near band edges, which have implications we expound in later sections on material and device characteristics such as mobility, subthreshold swing, and ON-OFF ratio.

\begin{figure}[h!]
\centering
\includegraphics*[width=.7\textwidth]{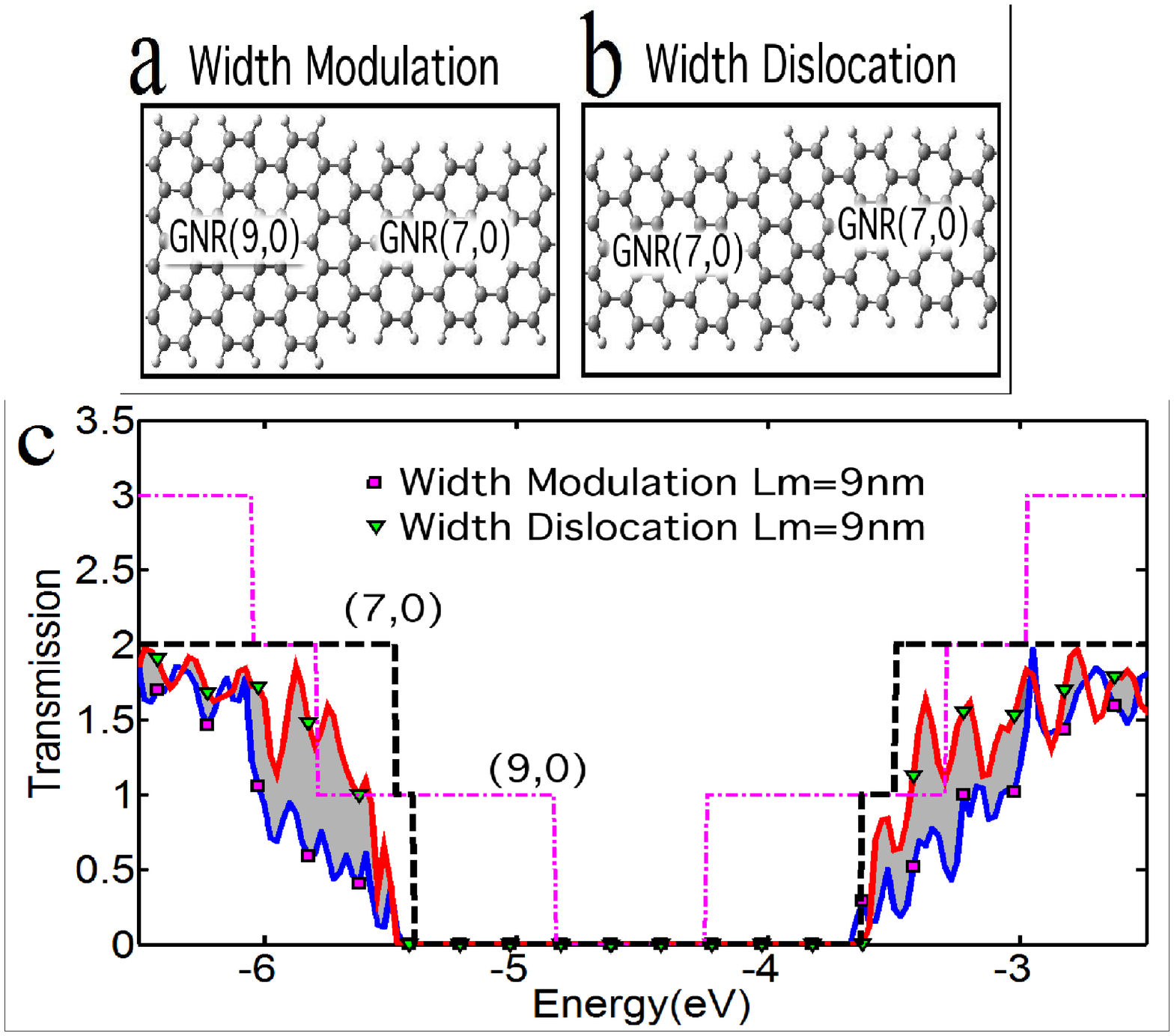}
\caption[]{(Top) Two kinds of roughness include variation in GNR width and a width dislocation across a slip line, maintaining the same width (Bottom) Transmission plots show that for either case, segments with the larger bandgap filter out the rest of the segments, thereby promoting the chiral curve with the highest bandgap, in agreement with experiments and EHT calculations ~\cite{tse01}.}
\label{ehtrough}       
\end{figure}

We have studied a wide spectrum of line edge roughnesses that can ultimately be classified as either a {\it{width modulation}} or {\it{width dislocation}}. Modulation in width along the armchair edge has a corresponding modulation in bandgap that follows an oscillating inverse square law relation between bandgap and width, spanning the three chiral curves. Meanwhile a width dislocation is an in-plane displacement in the GNR that energetically sees the same bandgap at interface of the slip dislocation, albeit with localized interfacial states. We introduced edge roughness in our geometries using a Gaussian white noise that adds or removes an integer number of dimer pairs with a correlation length $L_m$ ~\cite{are01}. The results of our EHT simulations with a statistics of rough edges explain why measured data are seen to cluster around one the $3p+1$ chirality curves.

An E-k relation cannot be rigorously defined for a structure without periodicity, so we focus instead on the transmission bandgap of the entire structure, calculated using the Non-Equilibrium Green's Function (NEGF) formalism in its simplest, Landauer level implementation.
The plotted transmission of the rough segment (Fig.~\ref{ehtrough}), sandwiched between two bulk metallic contacts, is shown for a random mixture of (7,0) and (9,0) chiralities ~\cite{tse01}. The rough GNR can be viewed as a random mixture of individual GNR segments with well defined chiralities and bandgaps, as long as the correlation length $L_m$ is larger than the electron wavelengths.
In the large majority of our simulations, we find that the transmission bandgap for the rough GNR follows the largest bandgap of the individual segments (in this case, the (7,0) segment). The larger bandgapped regions effectively filter out the incoming currents from the rest of the crystal, explaining the observed predilection towards the $3p+1$ chiralities. There are a few outliers however, corresponding to accidental situations where the segments with large bandgap are thin enough to become transparent tunneling. As we increase the frequency of roughness by bringing the correlation length below the electron wavelength, we expect to see more tunneling events. Such high frequency roughnesses, as well as inelastic scattering, will mix chiralities more thoroughly and are predicted to spread out the data more evenly over the three chiral curves rather than clustering them around the dominant (3p+1) segment.

The upshot of this analysis is that for reasonably smooth GNRs with $L_m > \lambda_F$, we expect the GNR width to be the sole arbiter of metallicity, so that wide GNRs are metallic while narrow ($< 10$ nm) GNRs are semiconducting ~\cite{dai02}. While the analyses leading to this conclusion assumed some simplifications (e.g. ignoring structural modifications due to back-bonding with substrates for instance), experiments seem to support this conclusion.

\section*{18.2. The next circle -- two terminal mobilities and I-Vs}
Our next circle would move on from the material parameters to electronic
properties such as its carrier density, mobility, conductivity, and
ultimately its current-voltage (I-V) characteristics.

\subsection*{18.2.1. Current-voltage characteristics (I-Vs)}
The Landauer expression gives us a convenient starting point for the current through any material,
\begin{equation}
I = \frac{2q}{h}\int  T(E)\, M(E) [f_1(E)-f_2(E)] dE
\label{eq:landauer}
\end{equation}
where $T \approx \lambda_{sc}/(\lambda_{sc} + L)$ is the quantum mechanical transmission per mode, that relates its scattering length $\lambda_{sc}$ with its length $L$. The number
of modes $M = \gamma_{eff}D(E)$, $D(E)$ being the density of states,
and the effective injection rate is given by
$1/\gamma_{eff} = 1/\gamma_1 + 1/\gamma_2 + 1/\gamma_{ch}$,
where $\gamma_{1,2}$ are the broadenings from the contacts, and $\gamma_{ch} = \hbar v(E)/L$ is the intrinsic transport rate in the channel. Assuming the contact broadenings
are large so that the rate limiting step is $\gamma_{ch}$, we can replace $\gamma_{eff} \approx \gamma_{ch}$.

The band dispersion of graphitic materials, ranging from epi-G to sG, BLG, CNTs and GNRs are all described by a universal formula ~\cite{dat01}
\begin{equation}
E = \pm \sqrt{E_{C,V}^2 + \hbar^2v_0^2k^2}
\end{equation}
where the band-edges are at $E_{C,V}$ while the high energy velocity in the linear regime is $v_0 \approx 10^8$ cm/s. From the dispersion, we can readily extract the 2D
density of states and band velocities
\begin{eqnarray}
D(E) &=& \Biggl(\frac{2WL}{\pi\hbar^2v_0^2}\Biggr)|E|\Biggl[\theta(E-E_C) + \theta(-E_V - E)\Biggr]\nonumber\\
v(E) &=& v_0\sqrt{1 - E_{C,V}^2/E^2}
\label{DV}
\end{eqnarray}
There is an additional energy dependence in the scattering length $\lambda_{sc}$. For ballistic channels, this is energy-independent, while for charge impurity and edge roughness scattering, $\lambda_{sc} \propto E$, while for acoustic phonon scattering, $\lambda_{sc} \propto 1/E$. The actual dependences are a bit more complicated, but
these are reasonable approximations to adopt.

The algebra becomes particularly simple if we ignore the energy-dependence of $\lambda_{sc}$. We can then do the Landauer integral, leading to
\begin{equation}
I = \frac{8q}{h}\Biggl(\frac{\lambda W}{\pi\hbar v_0L}\Biggr)I_0
\end{equation}
where the shape function $I_0$ depends on the current flow regime. Assuming we start
with an n-doped graphene with a bandgap, we get

\begin{equation}
I_0\textbf{}=
\begin{cases}
\frac{1}{2}\Biggl[
\mu_1\sqrt{\mu_1^2-E_C^2} - E_C^2\cosh^{-1}\Biggl(\frac{\mu_1}{E_C}\Biggr)-
\mu_2\sqrt{\mu_2^2-E_C^2} + E_C^2\cosh^{-1}\Biggl(\frac{\mu_2}{E_C}\Biggr)

\Biggr]
 \\ \quad \quad \quad \quad  \quad  \quad \quad \quad \quad  \quad     \quad \quad \quad \quad  \quad  \quad \quad \quad \quad  \quad       \text{if $qV_D < E_F - E_C$,}                   \\

\frac{1}{2}\Biggl[
\mu_1\sqrt{\mu_1^2-E_C^2} - E_C^2\cosh^{-1}\Biggl(\frac{\mu_1}{E_C}\Biggr)
\Biggr]
  \\ \quad \quad \quad \quad  \quad  \quad \quad \quad \quad  \quad     \quad \quad \quad \quad  \quad  \quad \quad \quad \quad  \quad \text{if $E_F-E_C < qV_D < E_F + E_V$,}\\
 \frac{1}{2}\Biggl[
\mu_1\sqrt{\mu_1^2-E_C^2} - E_C^2\cosh^{-1}\Biggl(\frac{\mu_1}{E_C}\Biggr)
\Biggr]- \frac{1}{2}\Biggl[
\mu_2\sqrt{\mu_2^2-E_V^2} - E_V^2\cosh^{-1}\Biggl(\frac{\mu_2}{E_V
}\Biggr)
\Biggr]
\\ \quad \quad \quad \quad  \quad  \quad \quad \quad \quad  \quad     \quad \quad \quad \quad  \quad  \quad \quad \quad \quad  \quad\text{if $qV_D > E_F + E_V$.}
\end{cases}
\end{equation}

where $\mu_1 = E_F$ and $\mu_2 = E_F - qV_D$.
The expressions can be further simplified. In the linear regime, the
current looks like
\begin{equation}
I_{linear} \approx 2G_0M\Biggl(\frac{v_0}{v_F}\Biggr)^2V_D
\end{equation}
where $G_0 = q^2/h$, the number of modes $M \approx 2W/(\lambda_F/2)$, and
the Fermi velocity $v_F = v_0\sqrt{1-E_C^2/E_F^2}$. The saturation current
\begin{equation}
I_{sat} \approx 4G_0M\Biggl(\frac{E_F}{2q}\Biggr)
\end{equation}
while the band-to-band tunneling current at high bias varies quadratically as
\begin{equation}
I_{BTB} \approx 4G_0M\Biggl(\frac{v_0}{v_F}\Biggr)V_D\Biggl(\frac{qV_D}{2E_F}\Biggr)
\end{equation}

\begin{figure}[t]
\includegraphics*[width=1\textwidth]{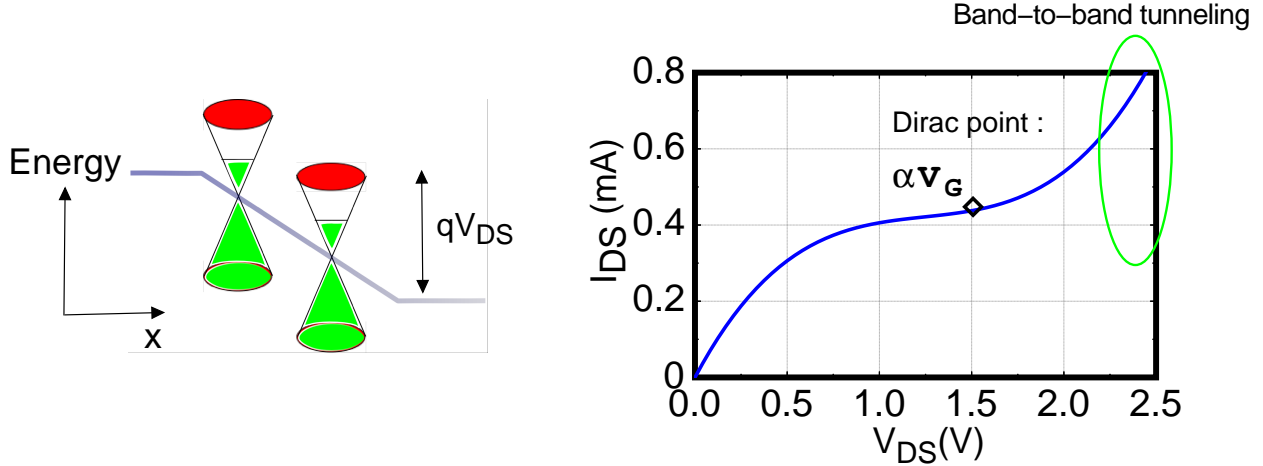}
\caption[]{A typical I-V output shows how the I-V tends to saturate at the Dirac point even without a bandgap. The shift in the Dirac point indicates the Laplace potential drop along the channel, eventually leading to band-to-band tunneling.}
\label{grdirac}
\end{figure}

\begin{figure}[t]
\includegraphics[width=0.48\textwidth]{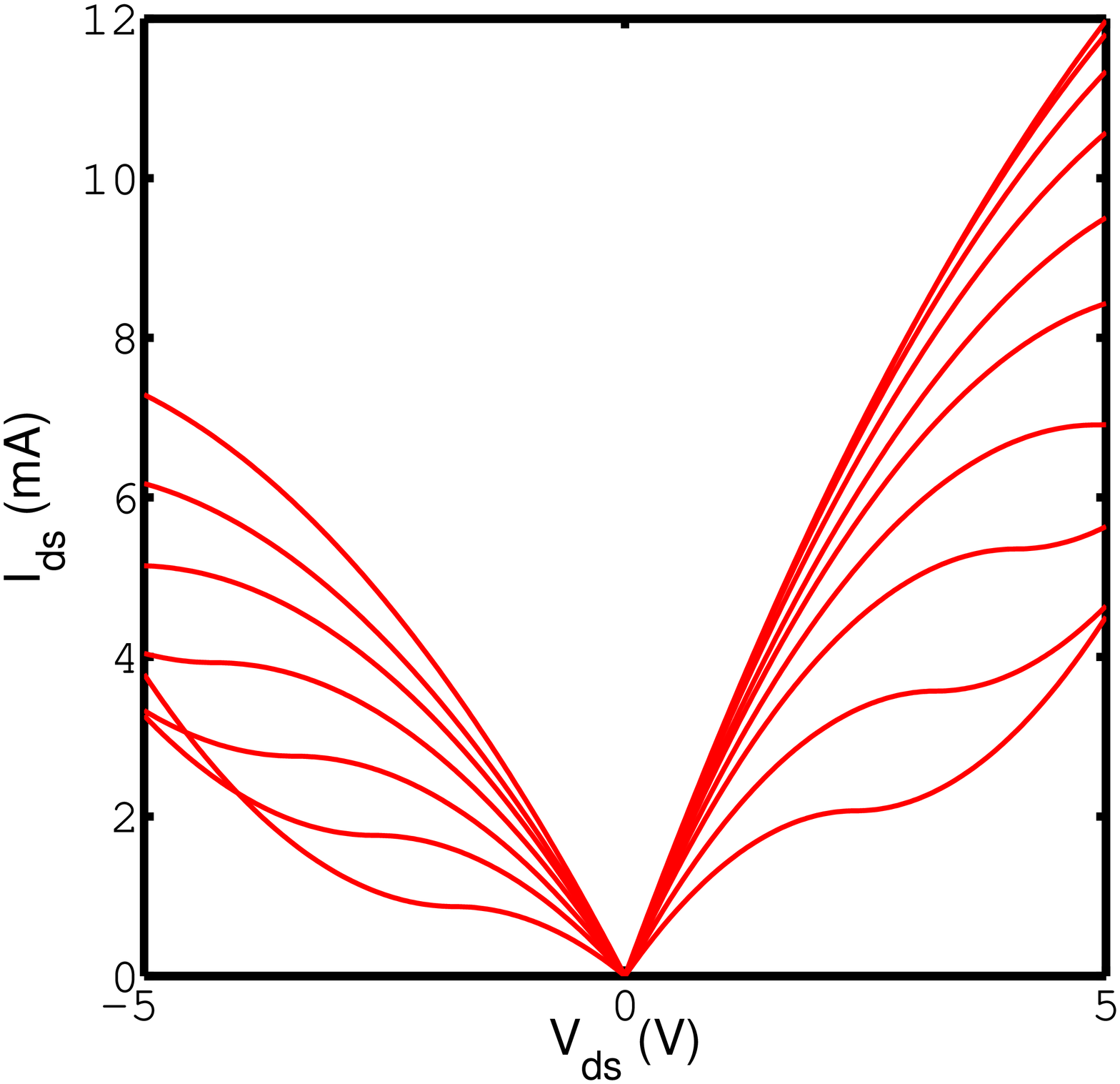}\label{a}
\includegraphics[width=0.5\textwidth]{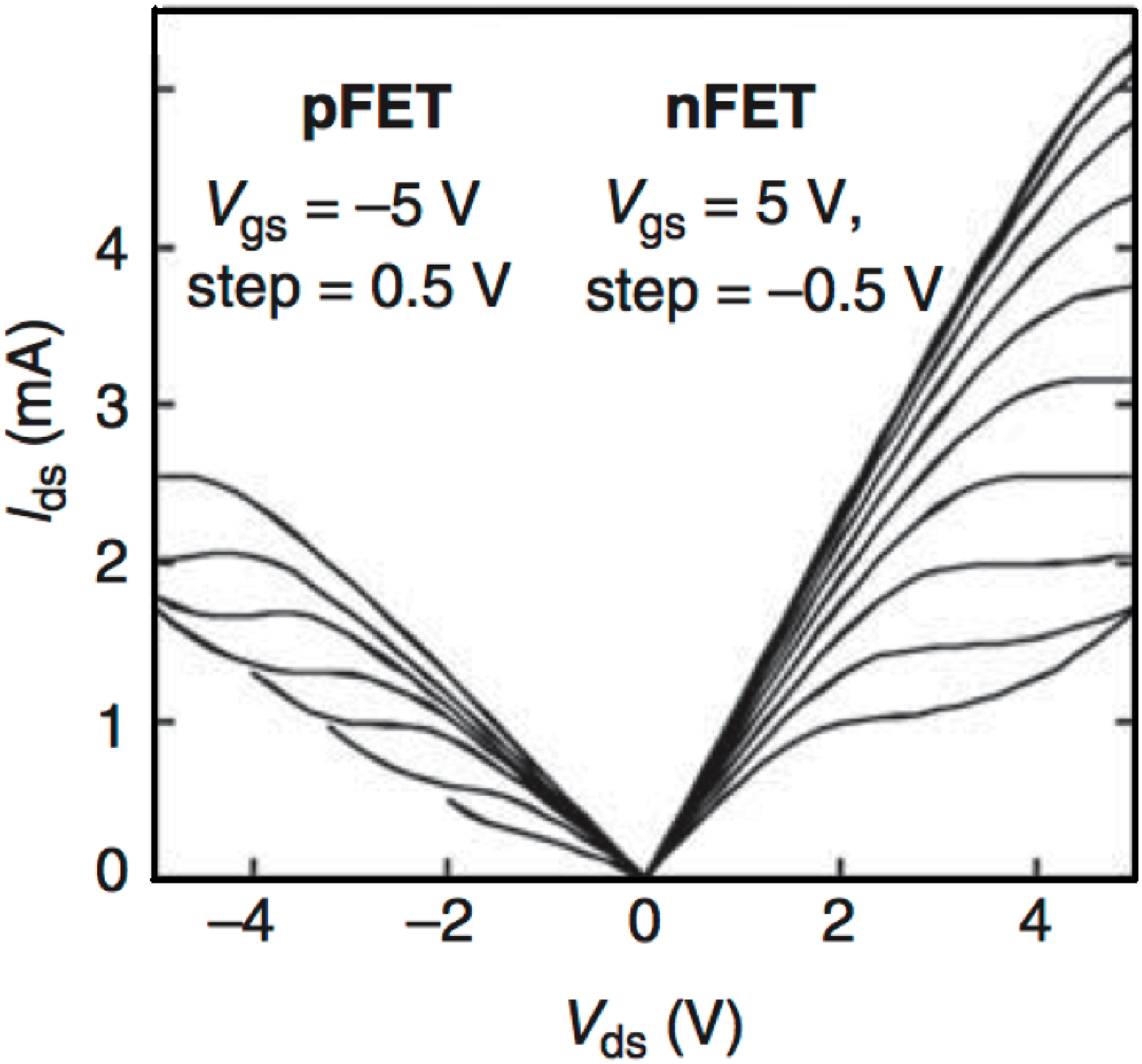}\label{b}
\caption[]{(Left) Theoretical and (Right) experimental I-Vs for graphene.
The calculations on the left assume a `doping' of the sheet by a charge
density that shifts the $K$-point relative to neutrality. We
also assume an inverse relation between the scattering length $\lambda_{sc}$
and the applied voltage on the n-side, consistent with scattering by
charge puddles associated with the above doping charge.
The data on the right are for graphene on SiC, where charge puddles and/or
back-bonding are expected to transfer a net charge density to the sheet ~\cite{moo01}.}
\label{fig:grasymm}
\end{figure}

Fig.~\ref{grdirac} shows typical I-Vs based on the above formula. These results agree with more involved, atomistic models for EHT coupled with non-equilibrium Green's function based simulations ~\cite{dat02}. The current shows a point of inflection at the Dirac point, which is shifted by the gate bias (bandgaps would give more extended saturating regions, as we will see for our three terminal I-Vs later on). The subsequent rise in current is indicative of band-to-band tunneling. Furthermore, a prominent I-V asymmetry, consistent with experiments on SiC, can be engineered into our I-Vs (Fig.~\ref{fig:grasymm}) readily by shifting the Fermi energy to simulate a charge transfer `doping~\cite{tsen4}' of 470meV through substrate impurities, back-bonding and/or charge puddle formation with SiC substrates. A mean-free-path($\lambda_{sc}$) that varied inversely with gate voltage was implemented in the left figure in Fig.~\ref{fig:grasymm}. For n-type conduction $\lambda_{sc}$ ranged between 18nm to 40nm and 20nm to 31nm for p-type. Typically we would expect at least 100nm for low bias conductance and down to 10nm as the biasing approaches the saturation and band-to-band tunneling regions. Chosen $\lambda_{sc}$ represent an average scattering length for the different regions. A more accurate model for scattering is necessary and will be developed in future works. In contrast, SiO$_2$ seems to dope the sheets minimally and the measured I-Vs show the expected symmetry between the electron and hole conducting sectors.

From the low-bias I-Vs, we can now extract the conductance $G = I_{linear}/V_D$, thence the sheet conductance $\sigma_s$ using $G = \sigma_sW/L$, and finally the mobility $\mu_s$ using $\sigma_s = qn_s\mu_s$, where $n_s$ is the sheet charge density related to the Fermi wavevector as $k_F = \sqrt{\pi n_s}$. Let us focus on the mobility first, keeping in mind that the effective mass $m^*$ in GNRs is {\it{energy-dependent}}.

\subsection*{18.2.2. Low bias mobility-bandgap tradeoffs: asymptotic band constraints}

Graphene's linear dispersion is known for contributing to an ultra-high mobility, but often overlooked is its origin in the low bandgap that ultimately hampers its ON-OFF ratio as an electronic switch. The carrier velocity ($v=1/\hbar dE/dk$) fundamentally saturates to $v_0 = {3a_0t}/{2\hbar} \sim 10^8 cm/s$, which forces its high energy bandstructure to a linear form regardless of bandgap size. Regardless of the mechanism of bandgap opening, or the particular progeny of graphene that we are looking at (epi-G, sG, CNT, GNR or BLG), the bandstructures are always writeable as
$E(k)\approx\pm\sqrt{(E_{G}/2)^{2} + (\hbar v_0 k)^{2}}$ ~\cite{rev01}.
Such an intimate relation between bandgap and dispersion connecting ultimately to its  conduction/valence band effective masses (as opposed to mid-gap tunneling effective mass) is unique in materials science.

\begin{figure}[h!]
\centering
\includegraphics*[width=1\textwidth]{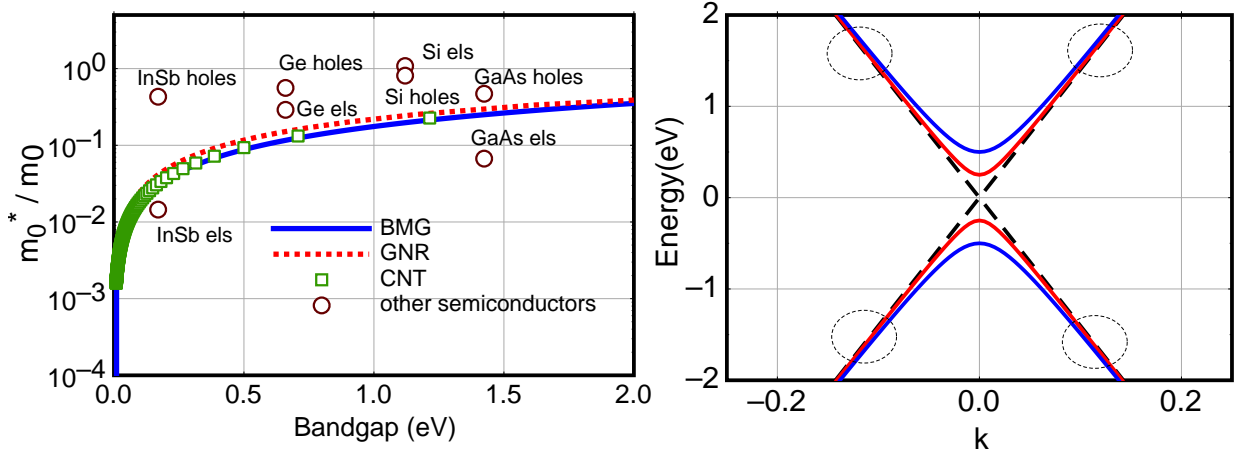}
\caption[]{Scaling of graphite effective masses shows that increasing the bandgap increases the mass $m^* = p_F/v_F$ due to the decrease in average curvature arising from a pinning of the E-k at high energy values ~\cite{sze01}.}
\label{grmass}       
\end{figure}

An extended bandgap constrained by the high energy velocity saturation localizes carriers, which shows up as a decrease in curvature at the band edges. It is easy to show from the above dispersion that the effective mass at each band-edge satisfies
\begin{equation}
\label{mobility1}
m_0^* = E_G/v_0^2
\end{equation}
indicating that the kinetic energy gained by the electrons and holes upon bandgap opening is taken ultimately from the corresponding crystal potential (Fig.~\ref{grmass}).

For transport considerations corresponding to high bias electrons and holes, we need to generalize the concept of effective mass to points away from the band-bottom, using $m^* = p/v = \hbar k/[1/\hbar(\partial E/\partial k)]$. This expression reduces to the usual dependence on curvature near the band-bottoms upon using L'Hospital's rule with $k \rightarrow 0$. In other words, effective mass and carrier velocity must be treated as energy dependent variables instead of material specific constants. From here, we can then extract the mobility $\mu=q\lambda_{sc} / p$, where carrier momentum $p=\hbar k$ and $\lambda_{sc}$ is the mean-free-path, related by $\lambda_{sc} = \pi/2(v_F\tau)$.
From the band dispersion or \emph{E versus k} relation we define the Fermi wavevector as:
\begin{eqnarray}
\label{k1}
k_F & = & \frac{ \sqrt{{E_F^{2}}-{{E_{gap}^{2}} /{4}}}}{ \hbar v_0 } \\
\label{k2}
             & = & \sqrt{\pi n_S}.
\end{eqnarray}
where  ${E_{gap}}/{2}$+$\alpha_{G}qV_{G}$  marks the position of the Fermi level ($E_F$) relative to the Dirac-point and $n_S$ is the electron density. Relating Eqs.~\ref{k1} and ~\ref{k2}, we can express a voltage and bandgap dependent electron density to determine the mobility
\begin{equation}
\label{mobility1}
\mu = \frac{q \tau}{m^*}=\frac{q \lambda_{sc}}{m^{*}v} = \frac{q\lambda_{sc}}{\hbar k_F}
\end{equation}

It is thus clear that the mobility depends on the value of $k_F$. As we vary the bandgap of graphitic systems (epi-G, BLG, s-G, or GNR), the variation in $k_F$ (equivalently, $E_F$) depends on what parameters are being held constant in the process. To start, let us assume the scattering length $\lambda$ is independent of energy, so that we're effectively working in the ballistic limit. At this point, we can assume the electron density $n_s$ is constant while the bandgap is being opened, so that $k_F$ is constant and the mobility does not change. However, possibly a more suitable metric is the gate overdrive $V_G - V_T$, which ultimately determines the charge density too using $n_s = C_G(V_G-V_T)$, where $C_G^{-1} = C_{ox}^{-1} + C_Q^{-1}$ involves both oxide and quantum  capacitances. In the limit of small density of states ($C_Q \ll C_{ox}$) at smaller bandgaps, the gate overdrive is the quantity that is controlled externally, and this changes $n_s$ as the quantum capacitance proportional to density of states increases with energy.
We then get
\begin{eqnarray}
\label{elc_den}
n_{S} = \frac{\alpha_{G}qV_G(\alpha_{G}qV_G + E_{gap})} {L\pi\hbar^2v_0^2} \\
\label{mobility2}
\mu =  \frac{q \lambda } {\hbar \sqrt{\pi n_{S}(\alpha_{G}qV_{G}, E_{gap})}}
\end{eqnarray}
where the gate transfer factor is $\alpha_{G}=C_{ox}/C_{\Sigma}$ and $C_{\Sigma}$ is the equivalent capacitance of a three-terminal device including its quantum capacitance. The fundamental mean-free-path ($\lambda_{sc}$) can be approximated semiclassically from the conductance $\sigma_s = q^2 D(E_F){\cal{D}} \approx \displaystyle\frac{2q^2}{h}\{\lambda_{sc}  k_F\}$ with $k_F$ defined in Eq.~\ref{k2} and ${\cal{D}}$ being the diffusion constant. Single layer graphene (SLG) can have $\lambda_{sc}$ on the order of microns
In addition to $\lambda_{sc}$, a more complete  of mean-free-path ($\lambda$) from Eq.~\ref{mobility2} would include scattering due to charged impurities, roughness, and possible phonons from interfacial materials given that phonons native to graphene are inherently suppressed, ($1/\lambda=1/\lambda_{sc} + 1/\lambda_{impurities} + 1/\lambda_{rough} +1/\lambda_{ph}$).

\begin{figure}[h!]
\centering
\includegraphics*[width=1\textwidth]{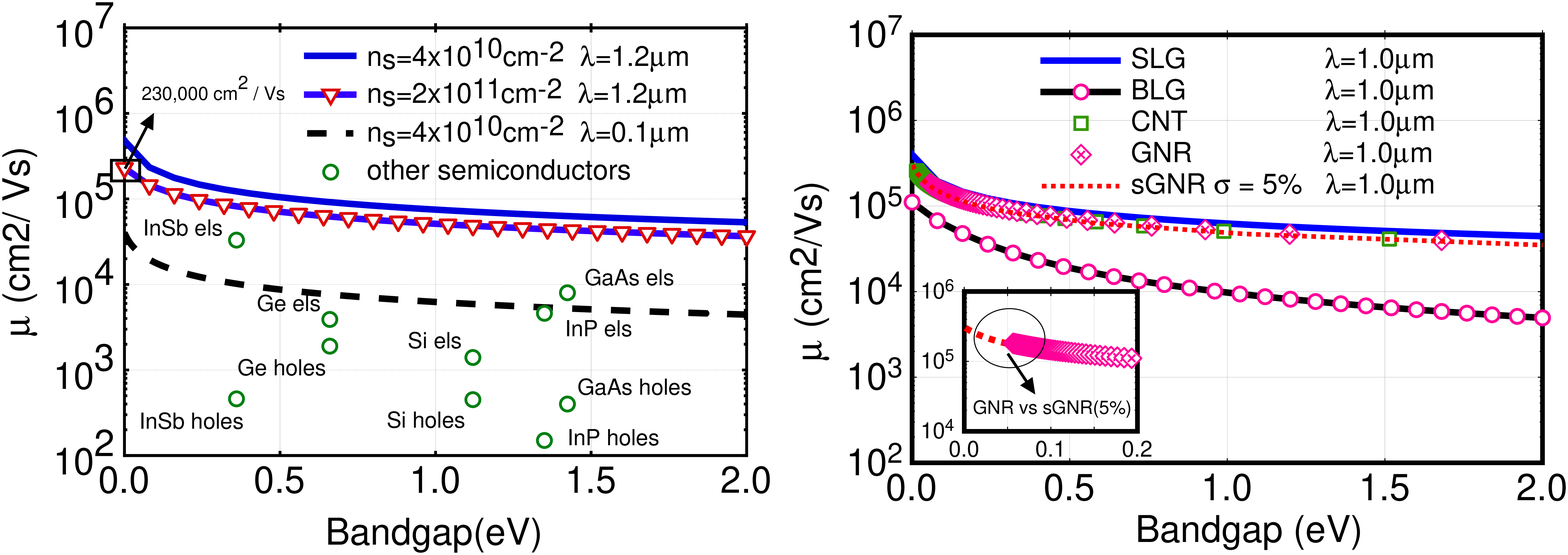}
\caption[]{Decrease in mobility (for fixed gate overdrive) between various graphitic materials as well as various electron densities }
\label{fig:5}       
\end{figure}

For a fixed gate overdrive, the mobility even for a ballistic device decreases
with bandgap, primarily due to the asymptotic constraint that pins the band
structure to a high energy linear dispersion. We emphasize that this trade-off
arises {\it{independent of any reduction in scattering length $\lambda_{sc}$ through
the bandgap opening process}}. The low effective mass of graphene arose from its
sharp conical bandstructure in the first place, so that opening a bandgap without removing the higher energy conical dispersion invariably makes the carriers heavier.

With respect to digital switching applications, the importance of the above trade-off
cannot be overstated. The mobility ultimately determines the switching speed through
the ON current, while the bandgap relates to the ON-OFF ratio. For cascaded devices,
it is also worth emphasizing that the ON-OFF ratio needs to be computed at high bias,
as the drain and gate terminals in regular CMOS like cascaded geometries are connected to the same supply voltage. Increasing the ON-OFF ratio by increasing the bandgap is predicted thereby to reduce the switching speed. We must therefore evaluate GNRs on this {\it{entire $\mu-E_G$ curve}} rather than at an isolated point on this 2D plot.
Eq.~\ref{mobility2} elegantly relates $E_{gap}$ and $\mu$ for various $\lambda$'s. Analyzing the three parameters $E_{gap}-\mu-\lambda$ simultaneously allows us to project the performance of graphene derivatives and compare against other common semiconductors as seen in Fig. ~\ref{fig:5}.

\section*{18.3. The third level -- active three-terminal electronics}
We now move outward to the next circle on the Y-chart, towards three
terminal active electronic devices. Our main focus will be on a class
of patterned device-interconnect hybrids, where we see certain notable
advantages mainly on the electrostatics and the contact barriers,
but challenges with the small bandgap show up as band-to-band tunneling and modest ON-OFF ratio.

\subsection*{18.3.1. Wide Narrow Wide (WNW) - All graphene devices}

The analyses from the previous sections set the platform for evaluating the
I-V characteristics of GNR devices in presence of a third gate terminal.
Our lesson from section 18.1 indicated that the GNR metallicity is primarily
set by its ribbon width, showing that one might be able to
monolithically pattern a wide-narrow-wide all graphene device that flows
seamlessly from metallic channels to semiconducting interconnects. Experiments
have in fact shown the ability to carve out GNRs using either chemistry or
nanoparticle mobilities that snip the sheets almost perfectly along their
C-C bonds. GNRs as thin as 1 nm with perfect edges have been manufactured
chemically ~\cite{cai01}. It is thus interesting to query what the device level
advantages of such a monolithically patterned GNR would be. We will later
discuss the circuit level ramifications.

\begin{figure}[ht]
\centering
\includegraphics*[width=.8\textwidth]{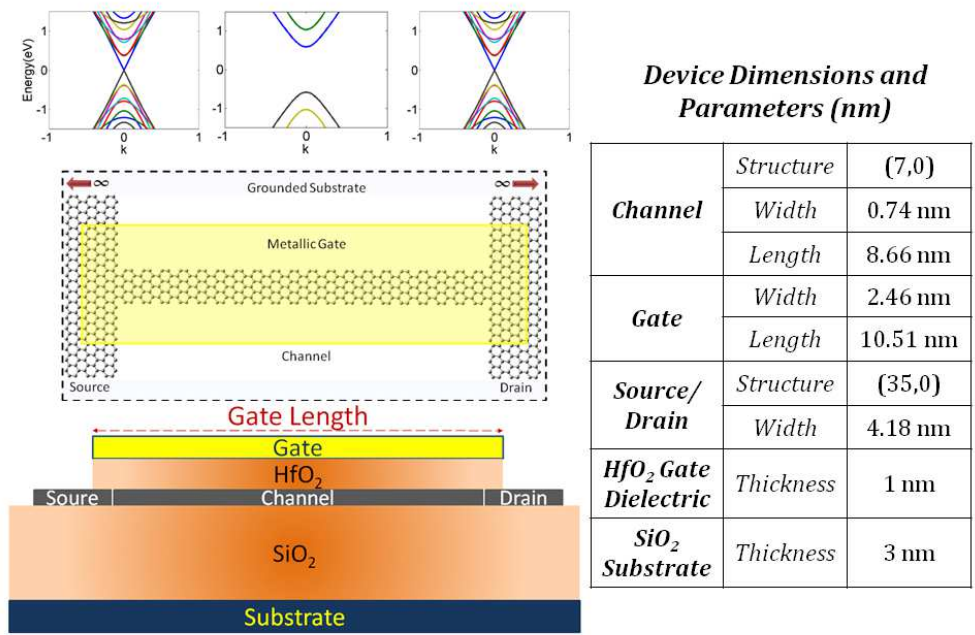}
\caption{WNW dual gated all graphene device, showing local $E-k$s (top), top view (center) and side view (bottom) with the device parameters listed.} \label{wnw}
\end{figure}

The structure of an imagined WNW graphene nanoribbon field-effect transistor (GNRFET) is shown in Fig.~\ref{wnw}. The wide regions
are metallic and the narrow ones semiconducting. There are planar gates
both at the top and the bottom, the top ones for gating and the bottom ones
for electrostatic `doping' (see figure later for inverters). Let us first discuss
how we simulate the I-V of one of these WNW devices.

\subsection*{18.3.2. Solving quantum transport and electrostatic equations}

The calculations we will show couples a suitable bandstructure/density of
states for the graphene channel with full 3D Poisson's equation for the
electrostatics and the Non-Equilibrium Green's Function (NEGF) formulation
for quantum transport ~\cite{dat01}. The wider contact regions are captured recursively
by computing their surface Green's functions $g_{1,2}(E)$. The corresponding energy-dependent
self-energy matrices $\Sigma_{1,2}(E) = \tau_{1,2}\, g_{1,2}\, \tau^\dagger_{1,2}$ project the contact states onto the
channel subspace, where the $\tau$ matrices capture the bonding between the contact and channel regions. In order to capture the interfacial chemistry properly, we
extend the device a couple of layers into the wider regions and calculated its Hamiltonian $H$ matrix. The Coulomb matrix $U$ is computed using the method-of-moments, described below ~\cite{ram01}.

From the above matrices, the retarded Green's function $G = (ES - H - U -\Sigma_1-\Sigma_2)^{-1}$ is computed, and thence the charge density matrix $\rho =
\int dE\, G \Sigma^{in} G^\dagger/2\pi$, whose trace gives us the total charge. $\Sigma^{in} = (\Gamma_1f_1+\Gamma_2f_2)$ in the simple limit where the only scattering arises at the contact channel interface. In the previous equation, $\Gamma_{1,2} = i(\Sigma_{1,2}-\Sigma^\dagger_{1,2})$ give the contact broadenings (the matrix analogues of the injection rates $\gamma_{1,2}$ introduced in section 18.2.1), while $f_{1,2}(E) = 1/[1+e^{(E-\mu_{1,2})/k_BT}]$ represent the contact Fermi-Dirac distributions, with $\mu_{1,2}$ being the bias-separated electrochemical potentials or quasi-Fermi energies in the contacts ~\cite{dat02}. The charge density matrix is then used to recompute the Coulomb matrix $U$ self-consistently through Poisson's equation.
Finally, the converged Green's function is used to compute the current $I = (2q/h)\int dE\, T(E) [f_1(E)-f_2(E)]$, where the transmission $T(E) = trace(\Gamma_1G\Gamma_2G^\dagger)$ ~\cite{dat01}.

Let us now get into a few details on the 3D Poisson equation we solve, using the method of moments (MOM) numerically. MOM captures the channel potential by setting up grid points on the individual device atoms with a specific charge density $\delta n_D$, and on the contact atoms with a specific applied voltage $\phi_C$ ~\cite{ram01}. Using the notations `C' for Contact and `D' for
Device, we get

\begin{equation}
\phi_d = \underbrace{(U_{dC}U_{CC}^{-1})\phi_C}_{Laplace} + \underbrace{(U_{dd}-U_{dC}U_{CC}^{-1}U_{Cd})}_{Single~Electron~Charging~Energy}\Delta n_d
\end{equation}
where we imply vector notations for the potentials $\phi$ and matrix notations for the Coulomb kernels $U$. $\Delta n_d$ is calculated relative to its neutrality value $N_0$ by tracing over $\rho$ above, while $N_0$ is calculated analogously, while grounding all the contact potentials
(this would depend on the workfunction of the contacts, as in MOS electrostatics).
The matrix elements in $U$ need to be computed with the correct dielectric constants.
Let us describe it in the simpler case with a dielectric constant $\kappa$ for the top gate and a dielectric constant unity for the bottom (trivially generalized to multiple dielectrics). Using the method of images,
\begin{eqnarray}
U(\vec{r}_1,\vec{r}_2) &=& \frac{q}{4\pi\epsilon_0\epsilon_1}\biggl[\frac{1}{|\vec{r}_1-\vec{r}_2|} - \biggl(\frac{\epsilon_2-\epsilon_1}
{\epsilon_2+\epsilon_1}\biggr)\frac{1}{|\vec{r}_1-\vec{r}^\prime_2|}\biggr]~~~({\rm{in~the~same~medium}})\nonumber\\
&=& \frac{q}{2\pi\epsilon_0(\epsilon_1+\epsilon_2)|\vec{r}_1-\vec{r}_2|}~~~({\rm{in~different~media}})
\end{eqnarray}
where $\vec{r}^\prime_2$ is the image of the charge at $\vec{r}_2$ ~\cite{jac01} ~\cite{neo01}.
A tricky point is to avoid the infinities at the onsite locations, for instance, when $x_1=x_2$ and $y_1=y_2$. We can avoid these using the Mataga-Nishimoto approximation,
where we replace terms like $1/|\vec{r}_1 - \vec{r}_2|$ with an atomistic correction
$1/\sqrt{|\vec{r}_1-\vec{r}_2|^2 + a^2}$, with the cut off parameter $a$ adjusted to represent the
correct onsite Coulomb (Hubbard) charging energy given by the difference between the atomic ionization
energy and the electron affinity ~\cite{mue01}.

Let us now discuss the observed electrostatic characteristics in the WNW device, which
explains the geometric advantages of this particular structure.

\subsection*{18.3.3. Improved electrostatics in 2-D}

\begin{figure}[h]
\centering
\includegraphics*[width=.9\textwidth]{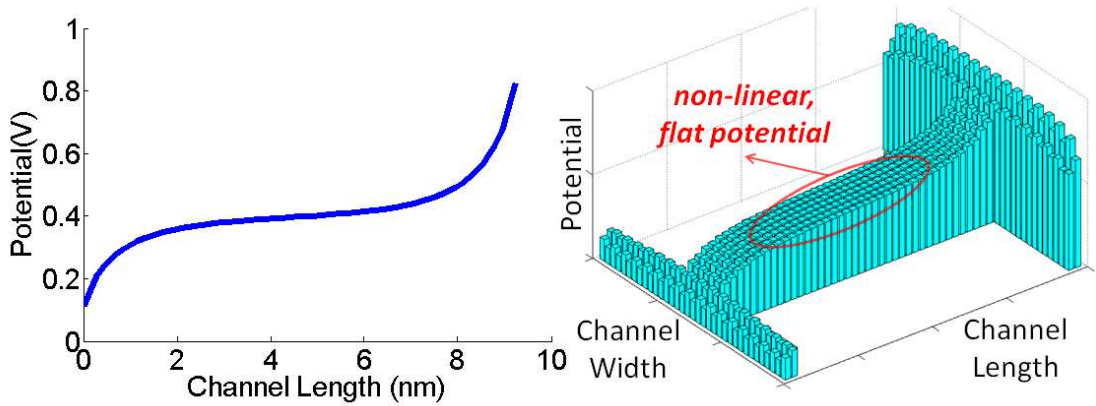}
\caption[]{(Left) The two-terminal potential shows the vanishing fields near the channel, implying the superior gate control and the improved short-channel effects with the 2D contacts. (Right) The 3D potential shows the non-linear flat potential in the middle of the channel.}
\label{fig:flat}       
\end{figure}

We simulate a device patterned monolithically from a two-dimensional sheet of graphene with a wide dilution of widths from the source and drain contacts to the active channel region. Simulated WNW (35-7-35) GNRFETs compose of (7,0) armchair graphene nanoribbon (GNR) narrow regions for the channel and (35,0) armchair GNR regions for the contact and interconnect regions. A metallic gate is placed on top of the channel region, while a wide grounded substrate is placed at the bottom of the channel. For calibration with the conventional CMOS technologies, the unique two-dimensional (2D) contacts of the GNRFET are replaced with 3D bulk metal contacts (whose surfaces act as parallel plate capacitors) for the same device, gate, and dielectric geometry.

A particular advantage of the WNW structure is the low capacitance of the 2D source drain contacts ~\cite{din01}. In a conventional MOSFET, the gate electrode needs to compete electrostatically with the source and drain for control of the channel charge. Indeed, a majority of developments in transistor technology over the last few decades have concentrated on making the field lines gate controlled rather than source/drain controlled. This is becoming harder with aggressive size scaling. The 2D source and drain contacts with a top gate makes the S/D capacitances lower, as they can only influence the channel through their fringing fields. Note that a 2D side gate geometry, as advocated in many device designs, would eliminate that electrostatic advantage, as the gate needs to compete with the S/D electrodes.

\begin{figure}[h]
\centering
\includegraphics*[width=.9\textwidth]{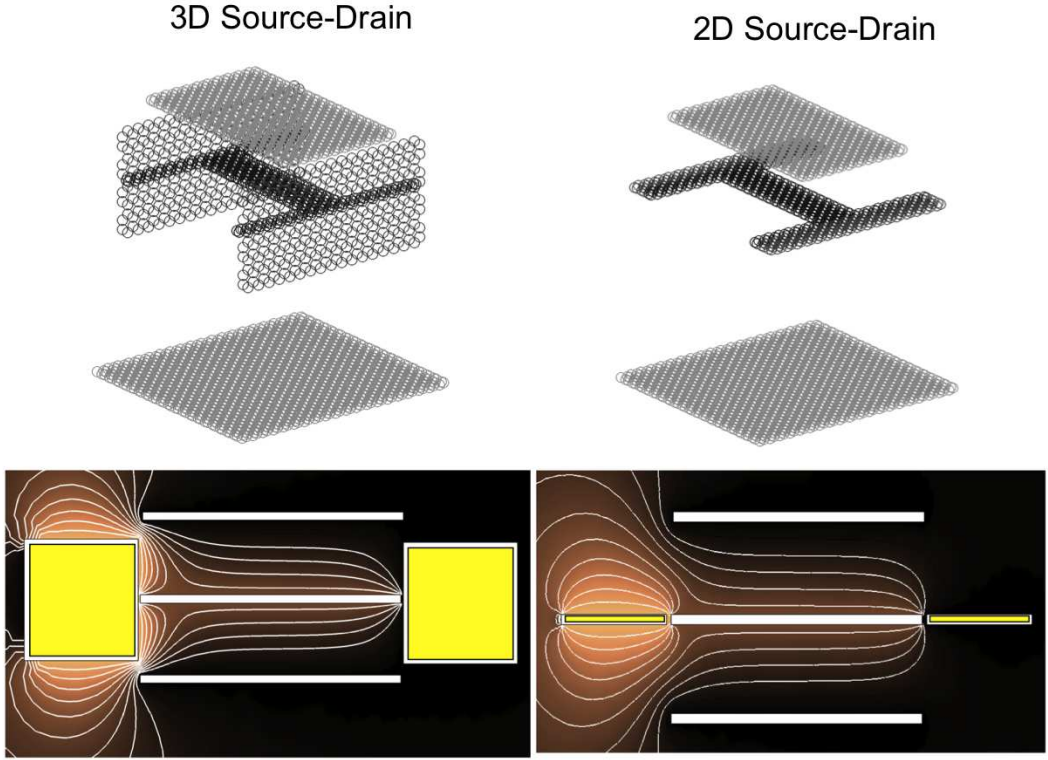}
\caption[]{Comparison of planar source drain
vs 3D source-drain. Denser field lines on the channel from the 3D contacts correlate to stronger source coupling and DIBL. For the given material and geometrical parameters listed in Fig.~\ref{wnw}, the $C_G/C_D$ ratios are 4.95 and 5.80 respectively. Top and bottom gates were grounded while the source was simulated with a potential of 0.3V and conducting channel had a potential of 0.1V}
\label{3D1D}       
\end{figure}

As the channel length gets shorter with the aggressively scaled technologies, the 3D contacts start to influence the channel potential as their surfaces act as parallel capacitor plates flanked by the insulator at the top and bottom. In the case of the monolithically patterned 2D GNR contacts, the charges on the contact surfaces are line charges so that the applied source-drain field decays into the channel, creating a {\it{non-linear channel potential even in the absence of a gate}} (Fig.~\ref{fig:flat}, left). Moving on to a three-terminal, dually gated structure, Fig.~\ref{fig:flat} (right), shows that the gate contact holds the channel potential flat against the action of the drain, thereby reducing short-channel effects.

\begin{figure}[h]
\centering
\includegraphics*[width=.9\textwidth]{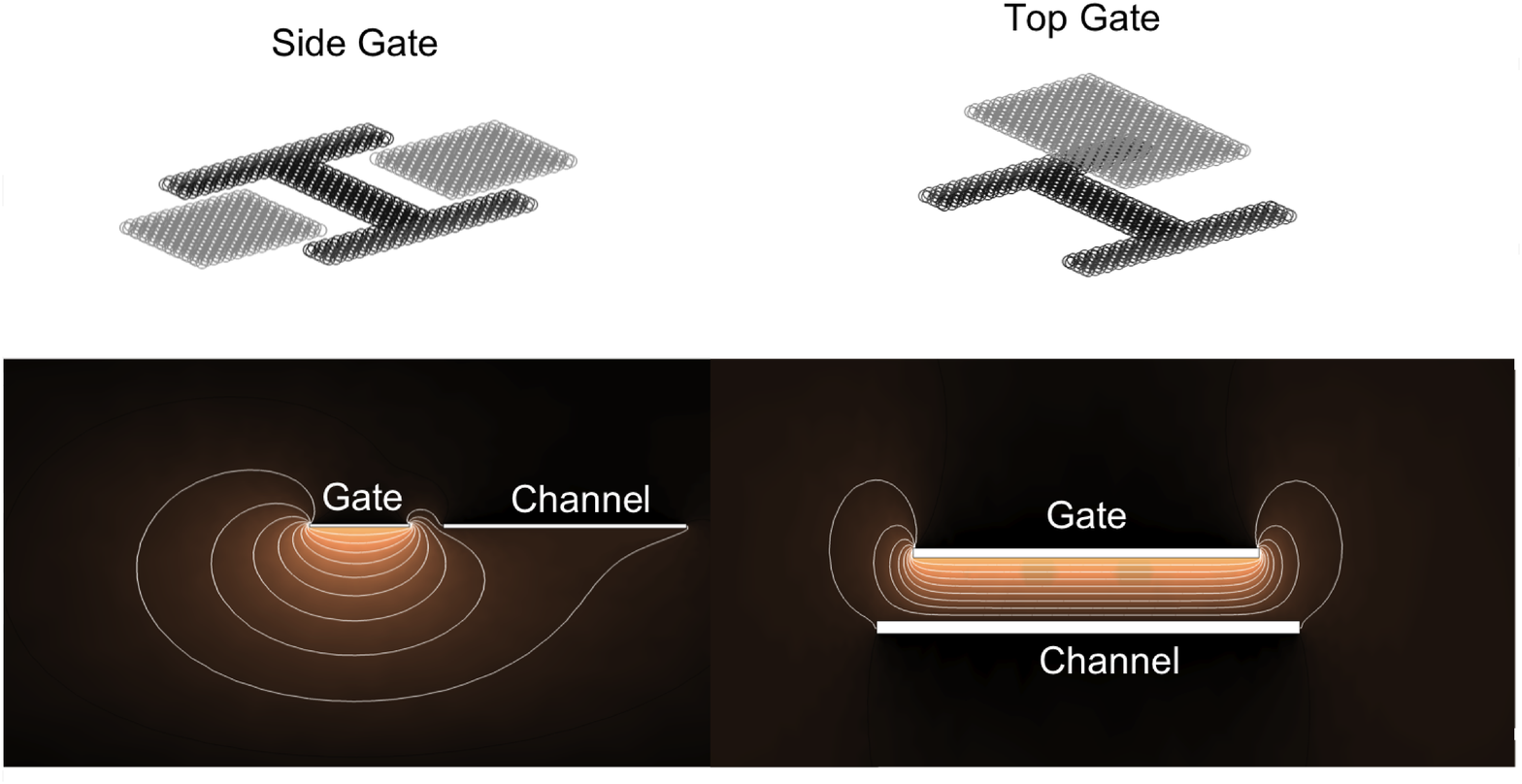}
\caption[]{Comparison of top vs side gate. Denser field lines from the top gate ensure better gate control which is reflected by larger gate capacitance. For the given material and geometrical parameters listed in Fig.~\ref{wnw}, the $C_G/C_D$ ratios are 5.80 and 3.82 respectively. Gates were biased at 0.4V , while the conducting channel had a potential of 0.1V. }
\label{topside}       
\end{figure}

Figs.~\ref{3D1D}, \ref{topside} show that for the same channel geometry, the top gate with 2D side contacts has the largest capacitance, followed by the top gate with 3D side contacts and finally the lowest gate to drain capacitance ratio is obtained when all electrodes are co-planar. The corresponding field line diagrams are also shown in these figures. Note also that in addition to the source, drain and dual gate electrodes, one needs to worry about the quantum capacitances, which are automatically included from our density matrix calculations that enter Poisson's equation.

The capacitance ratio can be extracted by plotting the channel transmission (T) for two scenarios: maintaining a constant drain voltage ($V_d$) while sweeping gate voltage ($V_g$) , and analogously, maintaining a constant $V_g$ while sweeping $V_d$. Sweeping the $V_g$ creates a larger energy shift in the transmission of the GNRFET channel than the sweeping of the $V_d$. From the shifting rates of these transmissions and the charge density calculations from the MOM, we can extract the capacitance values of the contacts. 
With shifts in transmission plots, we once again find that 2D contacts indeed help the gate exercise superior control over the channel ~\cite{din01}.

We will now explore the effect of the improved short channel effect on the computed I-V characteristics.

\subsection*{18.3.4. Three terminal I-Vs}

\begin{figure}[h]
\centering
\includegraphics*[width=.9\textwidth]{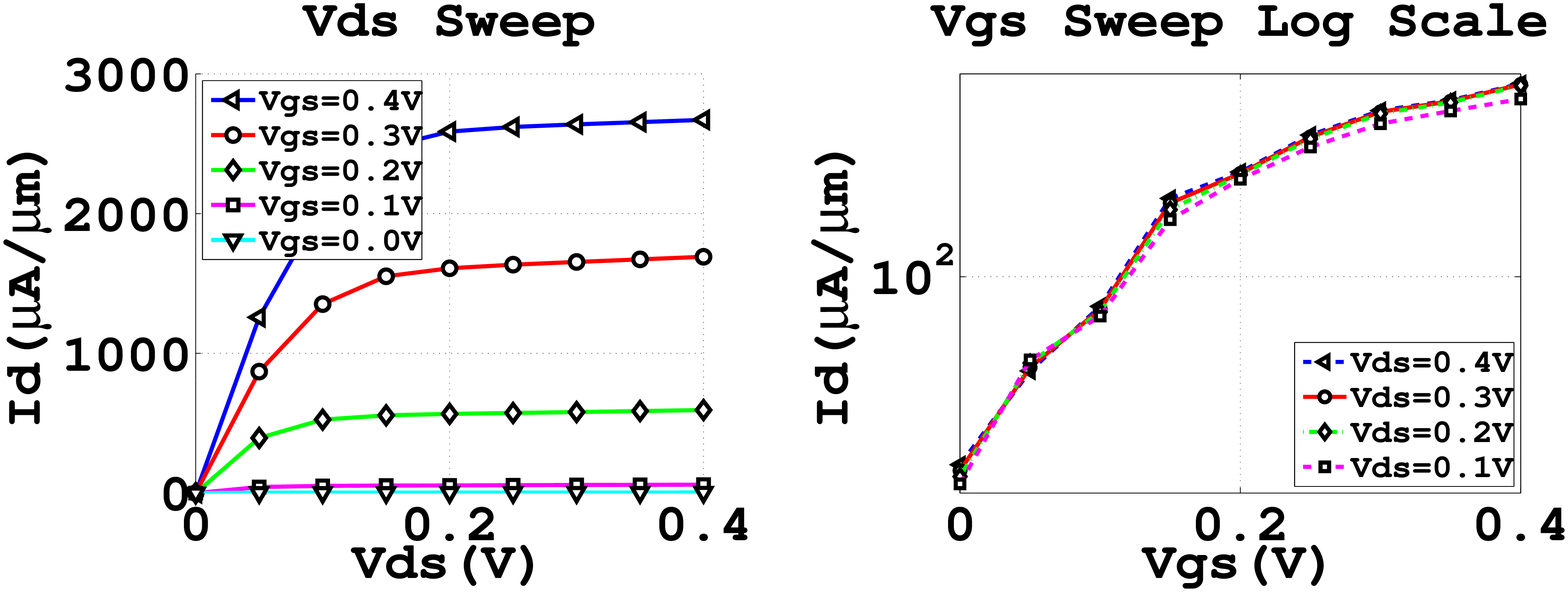}
\caption[]{I-V curves for a n-type GNRFET confined to create a large bandgap (in this case, a (7,0) armchair GNR with a bandgap nearly 1 eV). Such an extreme geometry postpones the onset of band-to-band tunneling. More importantly, the point of the I-V is to show the effect of better electrostatics which is independent of bandgap issues -- resulting in a high current saturation, low DIBL and SS.}
\label{fig:gnrfet_IV}       
\end{figure}

The computed  three terminal I-Vs (Fig.~\ref{fig:gnrfet_IV}) show excellent short channel effects, at least over a small voltage range given by the bandgap. Plotted vs gate voltage, the current shows excellent saturation characteristics with a large output impedance. Plotted vs gate voltage, the current shows little drain bias dependence (so-called drain induced barrier
lowering or DIBL). Taken together, the curves signify that the device electrostatics in the geometry is nearly ideal, making the outputs relatively robust with process variations.
It is interesting to note that instead of enhancing the gate capacitance as in regular CMOS devices, the trick in WNW devices has been to reduce the source and drain capacitances in comparison.

The simulations results of the model in the Fig.~\ref{fig:gnrfet_IV} demonstrate a Subthreshold Swing (SS) of ~84.3 mV/dec and a Drain-Induced Barrier Lowering (DIBL) of ~24 mV/V. We note that unless otherwise specified all simulations refer to material and geometrical parameters shown in Fig.~\ref{wnw} The value of the DIBL and the SS can be further improved by increasing the length of the channel (currently 1:8.6 ratio of HfO$_{2}$ thickness to channel length). These values calculated are better (smaller) than the estimated values of DIBL = 122 mV/V and the SS = 90 mV/dec for the double gate, 10 nm scaled Si MOSFETs ~\cite{has01}. Also in addition to showing improved short-channel effects, the GNRFET structure with the 2D contacts also shows controlled switching behavior. The on-current (Ion) of the system equals to ~2670.62 $A/\mu m$ with the off-current (Ioff) set at ~4.07 $A/\mu m$; thus giving a  Ion/Ioff ratio of ~656. The ON-OFF ratio, however, ends up being modest, and is a critical challenge in GNRFETs, especially in the light of its seemingly inverse relation with the charge mobility (section 18.2.2).

With the scaling of the channel length, the short-channel effects started to have a huge influence on the device parameters such as the DIBL and SS. As the channel lengths get shorter, the DIBL and SS of the device increases due to the $C_g$/$C_d$ ratio of decreasing with length. The line charges with the 2D contacts endow the gate with more control over the channel and interface states compared to the 3D contacts by lowering the drain capacitance. 

\subsection*{18.3.5. Pinning vs. Quasi-Ohmic contacts}

In today's semiconductors, Ohmic contacts are a desired to help achieve linear and asymmetric I-V characteristics. The potential profile inside the channel can be influenced by increasing the drain-source voltage ($V_{ds}$) or the gate voltage ($V_{gs}$).
For carbon nanotubes, this has been a particular challenge, as the metal carbon bonds
at the ends have predominantly created Schottky barriers ~\cite{guo01}. In our WNW geometries, since the bulk metal contacts are relegated to the ends of the device array, the bonding configuration near the wide-narrow interfaces are controlled by C-C covalent bonding. As our simulations show, this seems to promote a quasi-Ohmic behavior. The better bonding increases the decay lengths of the corresponding metal-induced gap states (MIGS) entering from the wide regions. The partial delocalization reduces the single-electron charging energy (that enters through our MOM treatment), thus making it harder for the contact regions to pin the Fermi energy and reducing the effectiveness of the Schottky barrier.

Schottky barrier FETs behave qualitatively different from MOSFETs 
In the latter, an applied gate bias reduces the channel potential and controls the thermionic emission over the voltage-dependent interfacial barrier. In the former, the gate reduces the thickness of the Schottky barrier and controls the tunneling of electrons through a voltage-independent,  pinned barrier height. The question is what the potential profile looks like in the channel, and whether the contact MIGS are effective in pinning this potential adequately.

\begin{figure}[h]
\centering
\includegraphics*[width=.9\textwidth]{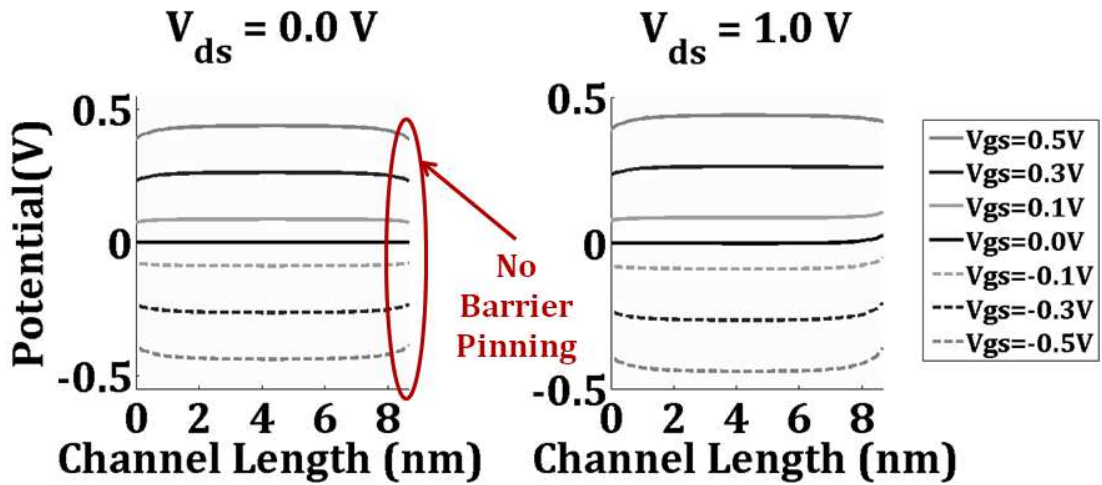}
\caption[]{At $V_{ds}$ = 0.0V and $V_{ds}$ = 1.0V, variation of channel potential with different gate voltages shows no barrier pinning at the contacts, implying Ohmic contacts.}
\label{fig:pinning}       
\end{figure}

As seen in the Fig.~\ref{fig:pinning}, the lowering of the potential throughout the entire graphene channel region with applied gate bias is a characteristic of the regular ohmic contact FETs rather than the Schottky barrier FETs, whose potentials would otherwise be pinned to the midgap by the charging of the interfacial states ~\cite{leo01}.

The MIGS due to the tail ends of the metal states in the contacts, leak in the semiconductor. Even with this 2D contact geometry, the MIGS will be present because of the contact-channel interfaces ~\cite{din01}. Our WNW all-graphene structure can filter these quickly decaying states, resulting in no significant contribution to the electron transmission. In the case of our device with the channel length of 8.66 nm, the MIGS do not travel all the way from source to the drain, but only extend approximately 0.7 nm into the semiconducting channel (Note a typo in one of our earlier papers, where we wrongly quoted this as 0.07 nm) ~\cite{din01}. The decay length of these MIGS can be calculated by plotting the wavefunction of the channel electrons at specific energies, 
as well as by evaluating the complex E-k diagram. The intensity of these MIGS at a given distance x can be calculated by using the equation $I_0$*$e^{(-x/2\lambda)}$, where  $I_0$ is the intensity of MIGS at the interface and $\lambda$ is the decay length.

Note that issues similar to those discussed here have been discussed in the context of pentacene molecules with CNT contacts. While CNTs would offer even better 1D electrostatic gains, a trade-off arises with the increasing series resistance in CNTs due to a paucity of modes. For GNR source/drain analogously, we will need to imagine wide blocks simultaneously contacting many GNR devices, so that the contact resistance is minimized by extending its width.

We now have all the tools to compute three-terminal I-Vs in graphitic structures, doing full justice to the complex electrostatics. Let us now see how this influences the circuit level performance metrics of GNRs.

\section*{18.4. The penultimate circle: GNR circuits}
Moving to the next level of abstraction and the last rung of our Gajski-Kuhn Y-chart, we focus on circuit level design issues when integrating GNR interconnects or devices. We start by discussing potential circuit topologies where we advocate a truly all graphene circuit, which is a logical extension to our WNW all graphene FET analogue. Next, instead of delving into an array of circuits we present fundamental circuit response design issues revolving around some basic GNR based logic element, namely the inverter. The inverter is the most basic logic circuit with the sole purpose of converting a logical 1(ON) to a logical 0(OFF) and vice versa for opposing input voltages. In the process of characterizing a GNR inverter we touch on interconnects and finally discuss drivability between connected GNR logic elements.

\begin{figure}[h]
\centering
\includegraphics*[width=0.9\textwidth]{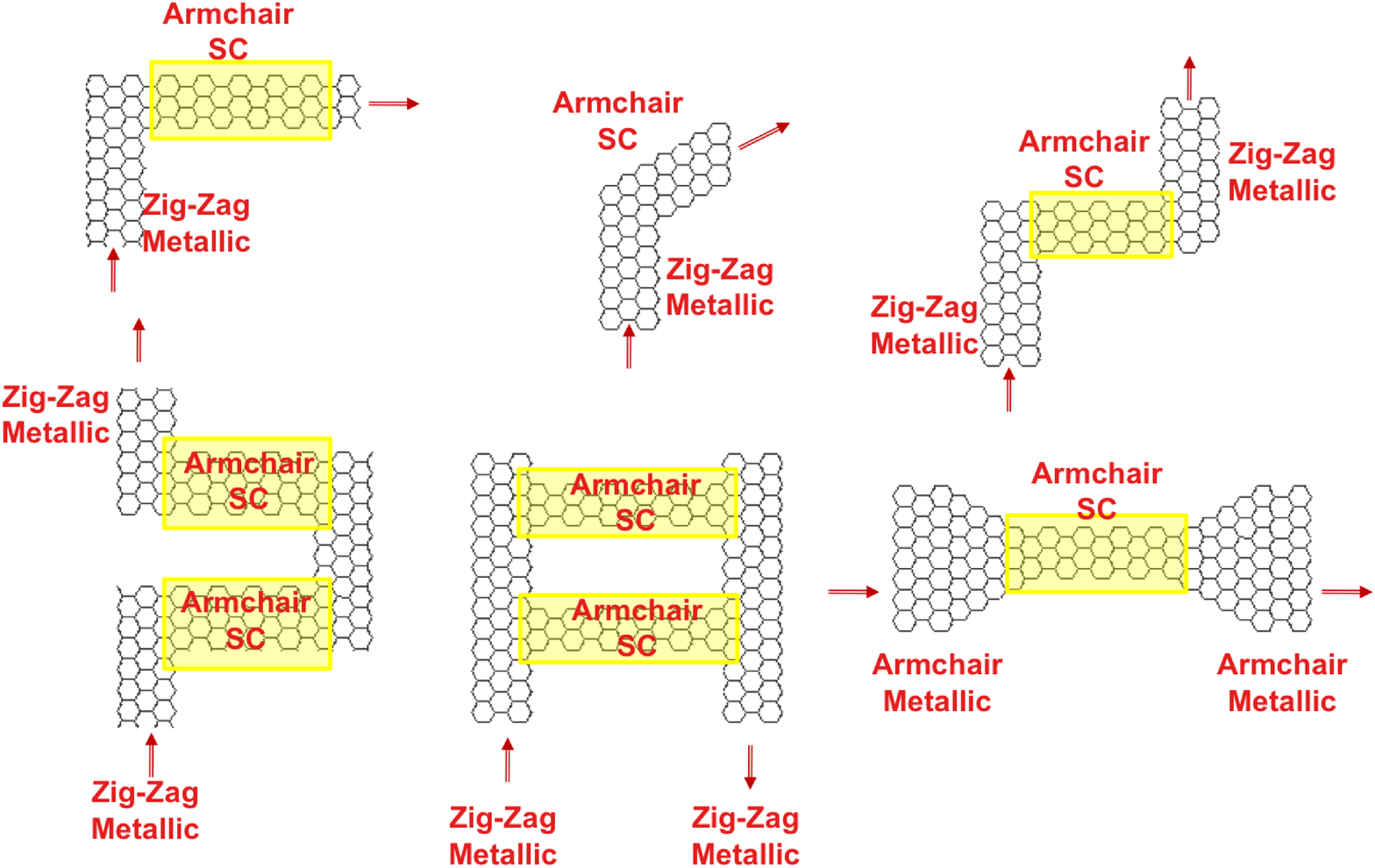}
\caption[]{The various building blocks for all graphene circuits}
\label{fig:blocks}       
\end{figure}

\subsection*{18.4.1. Geometry of an all graphene circuit}
An attractive feature of graphene for device engineers is its planarity, which stems from its $sp^2$ chemical bond hybridization.  Graphene's atomic flatness is compatible with existing lithographic device fabrication techniques established for CMOS.  Furthermore, exploiting how chirality in graphene influences electronic properties we present an array of building blocks for all graphene circuits (Fig.~\ref{fig:blocks}). In the previous circle on our Y-chart we explored an all armchair GNRFET. 

Circuit level enhancements start at the device level. Two important device performance metrics for digital circuit performance are ON-OFF current ratio and intrinsic gate propagation delay. GNRFET ON current scales proportionally with width, while the OFF current goes as $e^{Eg/kT}$ or $e^{c/WkT}$, where W is the width. To achieve manageable OFF currents for digital applications, GNR widths must be scaled within the sub-10nm regime to avoid increased in static power dissipation and poor ON-OFF ratio. However GNR scaling has little influence on propagation delay.

Propagation delay is defined as $ C_{IN} V_{dd} / I_{ON}$, where $C_{IN}$ is the intrinsic capacitance, $V_{dd}$ is the supply voltage and $I_{ON}$ is the saturating ON current at $V_{gs}$=$V_{dd}$. Increase in channel width increases $I_{ON}$ as threshold voltage decreases, which simultaneously drives up the OFF current exponentially. However $C_{IN}=(1/C_{ox} + 1/C_{Q})^{-1}$, which also scales with width balances any improvement in propagation delay from scaling $I_{ON}$ in graphene. From the perspective of a single GNR device there are not many options to improving the propagation delay. However a more useful context to address this issue would be on the circuit level where we have multiple logic elements. 

\begin{figure}[ht]
\includegraphics*[width=1\textwidth]{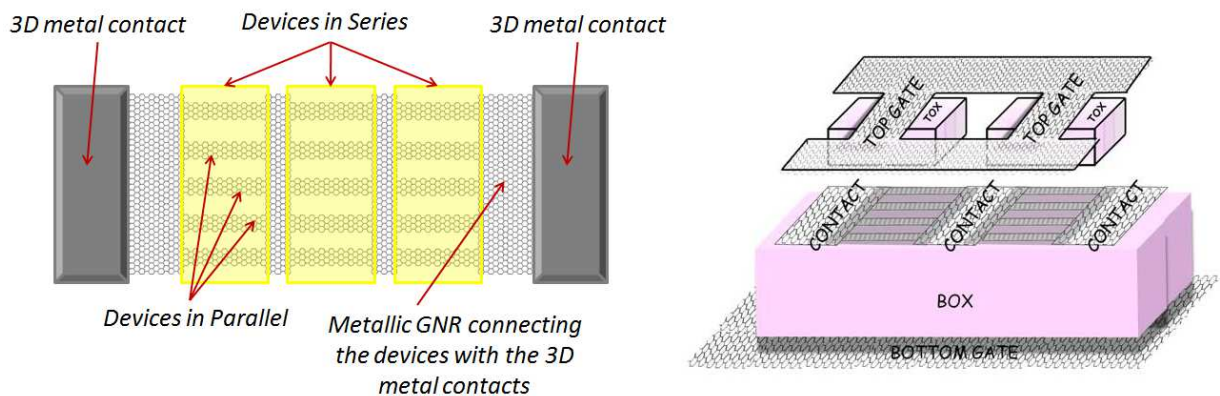}
\caption{Geometry of all graphene circuits} \label{fig:gnrckts}
\end{figure}

On a circuit level the complexity of propagation delay is in the load capacitance, which is now the sum of all capacitances at the output including the combinational drain capacitance from the previous inverter, interconnect capacitance($C_{interconnect}$), and next load gate capacitance. Using the building blocks in Fig.~\ref{fig:blocks} we present an all graphene circuit shown in Fig.~\ref{fig:gnrckts}, with cascaded GNRs in parallel on a semiconducting substrate with separate split gates for electrostatic doping the regions n and p-type. The cascaded GNRs could be separated by a high-k dielectric or even boron nitride in hexagonal lattice, which has the advantage of being atomically flat and  the absence of dangling bonds~\cite{tsen4} makes it less likely to carry adsorbents that could degrade the device. The advantage here is that ON-OFF ratio is held constant, while the increased ON current and capacitance which scales proportionally with the number of cascaded GNRs dilutes the parasitic interconnect capacitance, improving propagation delay and circuit performance.

Good quality graphene sheets have been made viable by current advancements in wafer-scale and pattern transfer techniques~\cite{tsen2,moo01}. However full realization of an all graphene circuit with various GNR interconnects and devices rely on the ability to pattern GNRs to narrow enough widths to produce a sizable bandgaps. Planar lithographic techniques are prone to edge roughness, while various chemical methods have had the most success in creating chemically precise GNR edges, but their applicability to scalable device level processes remain to be seen ~\cite{han01} ~\cite{fly01} ~\cite{dai01} ~\cite{kur01}. While roughness helps to make GNRs insensitive to chirality, we need further simulations to see how atomic fluctuations in the widths influence the corresponding threshold voltages and ON/OFF currents, an issue critical for the overall reliability of GNR circuits.

\subsection*{18.4.2. Compact Model Equations}

To simulate the performance of such a circuit, let us first outline a compact model. This will require us to outline (a) an equation for the bandstructure that includes effects due to edge strain and roughness, (b) an equation for the scattering length that depends on the phonon spectrum and edge roughness, (c) equations for the 2D electrostatics due to the source and drain contacts, and (d) the resulting I-Vs obtained by integrating the transmission over the relevant energy window.

To recap, the bandstructure of GNRs, including edge strain, can be written in a tight-binding form as $E = \pm \sqrt{E_{C,V}^2 + \hbar^2v_0^2k^2}$. Specific expressions for $E_{C,V}$ and $v_0$ for variously strained graphitic materials exist in Ref. ~\cite{gun01}.

The next term is the scattering $\lambda_{sc}$, which is related to the scattering time through an angle averaged geometrical factor and the overall Fermi velocity. The scattering time is extracted from Fermi's Golden Rule. For short range scattering by edge roughness and phonons, $\tau_{sc} \propto 1/|E|$, while for long ranged unscreened Coulomb scattering, $\tau_{sc} \propto |E|$. Explicit expressions exist in the literature ~\cite{sha01} ~\cite{fan01}.

The tricky part that does not exist in the literature are the electrostatic capacitances associated with the 2D electrostatics from the planar source and drain contacts, competing with the top and bottom gates through their individual dielectrics. We are in the process of extracting formulae based on knowledge of planar micro-strip line electrostatics, with geometrical factors calibrated with our numerical MOM solutions for a variety of geometries ~\cite{din01}.

Once we have the electrostatic, band and scattering parameters, we can then use Eq.~\ref{eq:landauer} to extract the I-Vs. For energy-independent $\lambda_{sc}$, this was already shown earlier. We will generalize it to various scattering configurations in our future work.

We thus have a comprehensive compact model that captures the chemistry and bandstructure, scattering, electrostatic and transport parameters needed for our circuit simulations. We will report one example here, and report further results in our subsequent publications.

\subsection*{18.4.3. Digital circuits:}

\begin{figure}[ht]
\includegraphics*[width=1\textwidth]{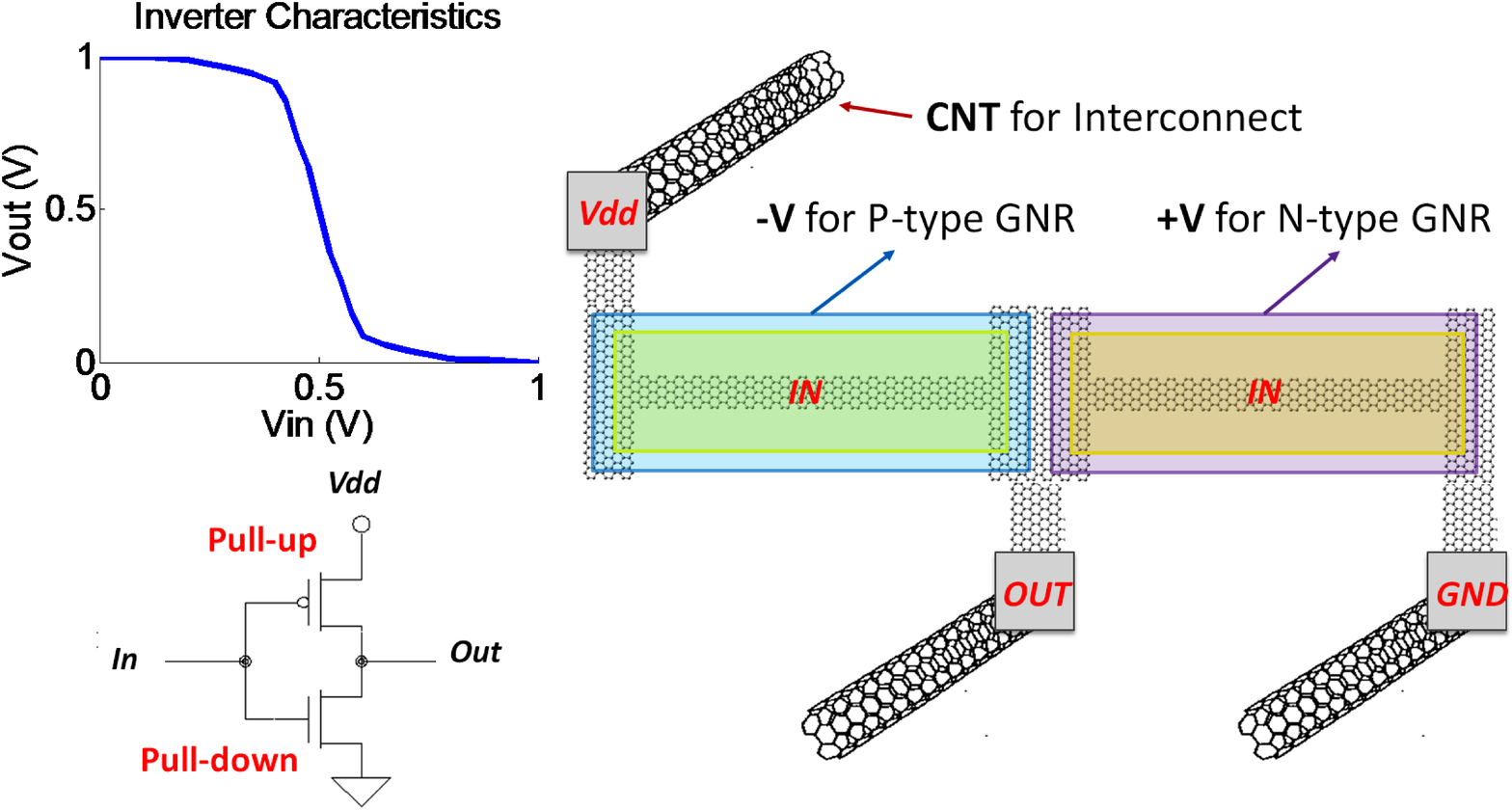}
\caption{GNR inverter geometry and voltage transfer curve. This inverter design uses the WNW (metal-semiconductor-metal) all graphene structure for  pull-up and pull-down networks. In this design CNT interconnects make direct contact with device level graphene. CNT/graphene interface has been experimentally demonstrated by Fujitsu Laboratories Ltd~\cite{tsen7,Fujitsu}.  }
 \label{fig:GNRINV}
\end{figure}

Static complementary CMOS gates utilize pull-up (PUN) and pull-down (PDN) networks to achieve low power dissipation and large noise margin in logic circuits such as the inverter, NAND, and NOR gates. CMOS logic circuits are composed of some series and parallel combinations of n and p-type FETs. An inverter illustrated in Fig.~\ref{fig:GNRINV} is the simplest logic element and the focus of this section of the review.

When the input into the common gate is Vin=0 the p-type FET (PUN) is active while the n-type FET (PDN) is cut-off, hence the circuit will pull the output voltage up toward the supply voltage (Vdd) or high, Vout=1. Likewise when Vin=1, n-type FET is active and p-type FET is cut-off pulling the circuit down toward ground, Vout=0. Usually it is impossible to pull-up or pull-down to exact values of 1 or 0, so threshold voltage and tolerance are designed for each circuit to help distinguishing between these two logic levels. Circuit designers allow some tolerance in the voltage levels used to avoid conditions that generate intermediate levels that are undefined. For example, 0 to 0.2V on the output can represent logic (0) and 0.3 to 0.5V can show (1), making the 0.2 to 0.3V range invalid, not metastable, since the circuits cannot instantly change voltage levels.  

\begin{figure}[ht]
\includegraphics*[width=1\textwidth]{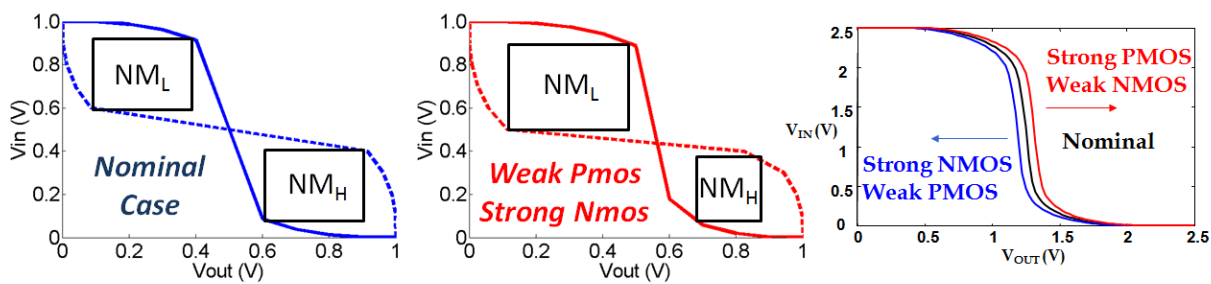}
\caption{showing the importance of balancing CMOS transistor sizes to achieve equal high and low noise margins(NM). The noise margin is graphically represented by the largest square that fits inside the enclosed space outlined by normal and rotated VTCs.}
\label{fig:PDNPUN}
\end{figure}

The voltage-transfer curve (VTC) of an inverter circuit captures the DC or steady-state of specific input versus output voltages and provides a figure of merit for the static behavior of the inverter. 
VTCs for logic circuits provide information on operating logic-levels at the output, noise margins, and gain. Ideally we want the VTC to appear as an inverted step-function, indicating precise switching between the on and off stages, but in real devices there is a continuous transition between on and off states.  From the VTC we can extract noise margin (Fig.~\ref{fig:PDNPUN}), which provides a measure of circuit reliability and predictability. Biasing outside the noise margin puts the logic circuit in an unpredictable state. Circuit designers want to maximize the noise margins.

\emph{A significant advantage of graphene is its intrinsic electron-hole effective mass symmetry.} In the absence of extrinsic doping a graphene based FET the I-V characteristics for n and p-type conduction would be the symmetric. However, asymmetry can be introduced into the system through charge-transfer doping~\cite{tsen4}
(e.g. Fig.~\ref{fig:grasymm}) or by contact induced doping~\cite{tsen5}. Significant screening of charge impurities in the substrate should bring Fermi level closer to its intrinsic value at the Dirac or $K-$point, therefore recovering symmetric n and p-type I-V characteristics. \emph{On a circuit level this symmetry means the response of PUN and PDN would be equal and opposite, which is important for circuit reliability, and not to mention ease of circuit design.} In conventional Si-CMOS logic circuits, the asymmetry in the electron-hole effective mass is compensated by scaling the physical width of the p-type FETs in the PUN so the I-Vs are equal and opposite with the PDN. Graphene's natural electron-hole symmetry would allow circuit designers to bypass this design issue.

\begin{figure}[h]
\includegraphics*[width=1\textwidth]{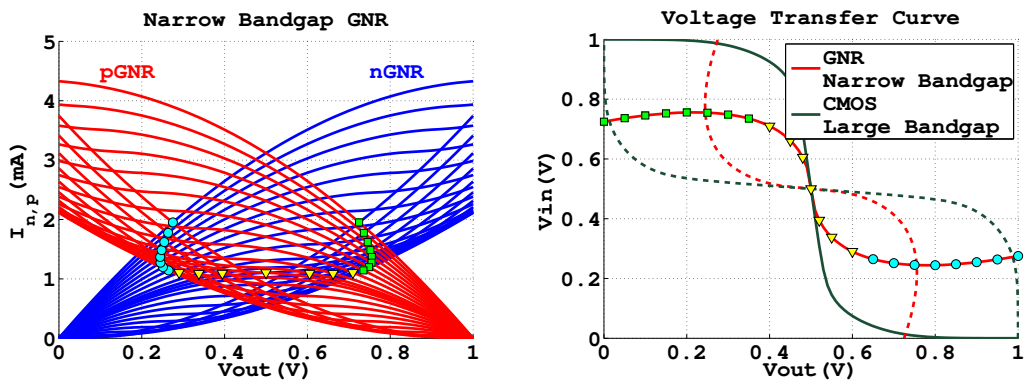}
\caption{Comparison of VTC curves for narrow bandgap GNR and 45nm CMOS
Technology. Narrow bandgap GNRFETs will be more susceptible to noise than
CMOS due to smaller noise margins.}
\label{fig:VTCC}
\end{figure}

A major impediment to GNR based logic circuits is its narrow bandgap ( $\leqslant 200meV$), as the device elements in the PUN and PDN are prone to sub-threshold leakage from band-to-band tunneling. The two-fold effect on an GNRFET-based inverter where the channel has a narrow bandgap is demonstrated in Fig.~\ref{fig:VTCC}. The first effect is a large voltage swing of approximately 0.4V. The second effect is a significantly diminished noise margin. Band-to-band tunneling in narrow bandgap GNRFETs prevents either the PUN or the PDN from completely cutting off when its complement network is active.

Fig.~\ref{fig:GNRINV} shows the physical layout of a functional graphene inverter composed of WNW P-type and N-type GNR device arrays and the voltage transfer curve. The inverter voltage-transfer curve and gain can be calculated readily from the current-voltage characteristics. As expected the gain of the device determined by the electrostatics, geometrical parameters, and mobilities which ultimately determine the P and N-type GNR transconductances. The VTC above with gain of 4 is derived from the I-V shown in Fig.~\ref{fig:GNRINV} for the 8.66nm device by using the methodology described in detail in ~\cite{cha03}. These I-Vs generated in SPICE can be used to simulate other complex layouts such as NAND or NOR gates shown in Fig.~\ref{fig:gnrnand} (The results of these logic gates will be reported in future publications).  

Propagation delay can be measured by pulsing the input voltage between 0 and 1 and observing the output transient response. The transit time for a GNRFET is approximately $L/v$, where $L$ is the length of the channel and $v$ is an energy dependent velocity defined in Eq.~\ref{DV}. Intrinsic and extrinsic device level scattering mechanics could also influence transit time. However, a cascade of inverters or some other logic elements in series, the load capacitance between each logic stage typically dominated by $C_{interconnect}$ would be responsible for the majority of the delay. The GNR circuit layout we presented earlier and show in Fig.~\ref{fig:gnrckts} takes advantage of non-scaling parasitic interconnect capacitances, lowering delay and improving performance. 

Beyond individual logical elements (ie., inverter, NAND, NOR), an important CMOS circuit design parameter is fan-out, which estimates the number of logic stages or CMOS gates that can be consecutively driven before signal attenuation is no longer tolerable. Past the maximum fan-out a repeater or amplifier is necessary to drive subsequent logic stages in a circuit. The maximum fan-out scales proportionally with propagation delay; therefore circuits designed for low frequency applications have a larger maximum fan-out compared to circuits designed for higher frequency applications. If graphene is to indeed follow the MOSFET and CMOS paradigm fan-out would be important circuit design trade-off to  consider and a topic we will discuss in an upcoming work.

\subsection*{18.4.4. Physical domain issues: monolithic device-interconnect structures}
\begin{figure}[h!]
\vspace*{0.5in}
\includegraphics*[width=1\textwidth]{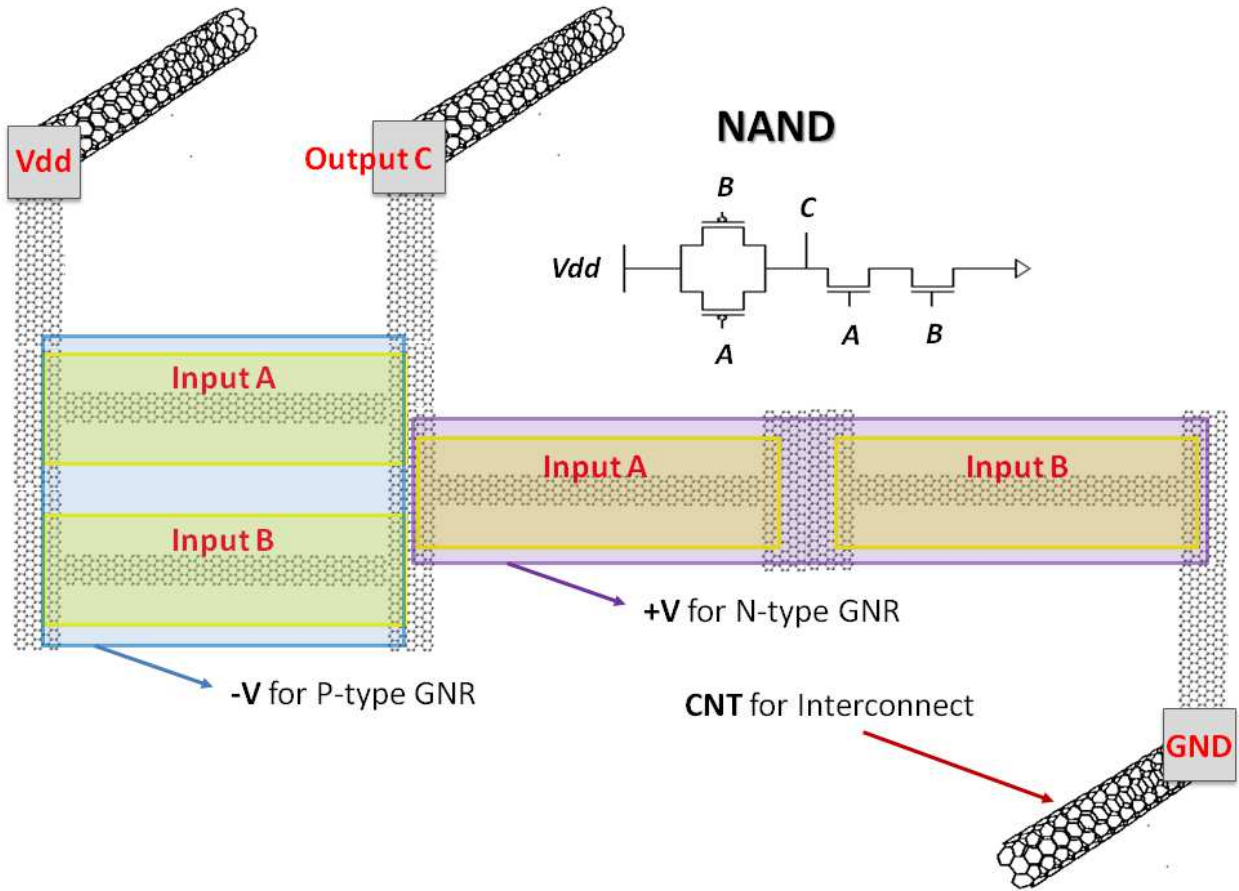}

\caption{GNR NAND layouts with electrostatic doped back-gates and
interconnecting `vias' between multiple levels. CNT/graphene interface has been experimentally demonstrated by Fujitsu Laboratories Ltd~\cite{tsen7,Fujitsu}.}
 \label{fig:gnrnand}
\end{figure}

The bane of many CNT-based circuit ideas is the degradation due to the dominant aspect of the Schottky contacts between the devices and interconnect. Luckily, GNR circuits can avoid this problem by using the same sheet of graphene for both active devices and interconnect, as seen in Fig.~\ref{fig:gnrnand}

Thus, for such monolithic GNR device/interconnect structures with Ohmic contacts, the behavior can be close to the ideal device predicted by simulations. Of course, contacts to metal are still required, but at a coarser granularity, thus the expected degradation in performance will be proportionally smaller. Contacts to metal are necessary for several reasons; first, the FET gates and drain/source electrodes cannot share the same graphene sheet, so connecting the output (drain) of a device to the input (gate) of the next will require a contact, second, topological requirements for connecting complex circuits can rarely be mapped to a planar graph, and non-planar graphs implicitly require more than one layer of interconnect, thus contacts, and for a really large circuit a large number of interconnect layers ($\sim$ 10 for a modern CMOS circuit).

\section*{18.5. Conclusions}

The future of microelectronics relies on continually scaling the critical dimensions of bulk CMOS technologies.
The semiconductor industry faces serious challenges in this respect, due to a host of technical and economic
constraints. One way to mitigate this is to use novel architectures, combining inherently scalable top-down
techniques, such as lithography, with novel bottom-up fabrication approaches, such as self-assembly. An alternate
way is to look for novel channel materials beyond silicon - a strong candidate being another group IV element, carbon, graphene being one of its distinguished allotropes.

Until recently, much attention has been focused on CNTs.
CNTs have already demonstrated excellent intrinsic performance, high gain, high
carrier mobility, high reliability, and are almost ideal devices in themselves ~\cite{bac01} ~\cite{mar01} ~\cite{tan01}. Unfortunately, there
seem to be no practical, scalable solutions for arranging multiple CNTs on a substrate uniformly with a small
pitch, needed to deliver adequate current for fast switching ~\cite{dai01} ~\cite{cas01}. Neither are there clear approaches for contacting
CNTs to interconnects or to each other at low impedance to realize complex circuits using large scale fabrication
schemes. In fact, CNTs frequently encounter Schottky barriers at the contacts that control the
tunneling electrons, hampering their reliability ~\cite{guo02}.

In contrast, graphene's planar profile makes it amenable to well-established planar fabrication techniques
for silicon CMOS devices. Its mobility can reach up to
two orders of magnitude above silicon, and the ability to engineer its bandgap with width alone points to the
feasibility of all-graphene devices that can exploit covalent bonding chemistry at the contact and better
inherent electrostatics to allow more traditional, MOSFET like gate control mechanisms. However its bane seems to
be its metallicity. As we saw earlier, graphene is naturally a zero-band gap
semi-metal, and attempts to open a bandgap, such as through strain, quantization or field asymmetries have limited
the bandgaps to $< 200$ mV ~\cite{cai01}. In a regular field effect transistor application, such a bandgap
translates to an ON-OFF ratio of $\sim$ 70, inadequate for digital logic ~\cite{lia01}. This seems consistent with experiments on high speed graphene transistors, which show I-Vs that are essentially quasi-linear. There is a lot of activity also on opening bandgaps in graphene, although it seems that this may reduce the mobility that made graphene a promising electronic device in the first place.

Despite the significant
challenges, graphene's high carrier mobility and fast switching speeds make it widely studied. This includes revolutionary applications such as based on charge focusing, graphene spintronics and excitonic condensation of pseudospin states ~\cite{red01}.
At the same time, there is wide activity on graphene based conventional electronic devices, specifically, RF devices and CMOS switches. It is not clear what the prospects of GNRs are vis-a-vis switching, given its low bandgap and its seemingly fundamental mobility-bandgap tradeoff. However, as we have argued here, there still are a few notable advantages to using graphene geometries, namely, (i) {\it{a natural electron-hole symmetry}} that helps with inverter design; (ii) {\it{its convenient placement between 1D and 3D}}, so that it offers distinct {electrostatic advantages without picking up too much series resistance arising from a paucity of modes; (iii) the {\it{tunability of its electronic properties}} primarily by controlling its width. Experiments are still emerging on these fronts, and it remains to be seen whether one can put these advantages to good use.

In this chapter, we made a first attempt to transverse the Gajski-Kuhn Y chart adapted to graphene, spanning physical, structural and behavioral domains for graphene. We ended with monolithic all graphene circuits and device/interconnect geometries, ultimately stopping short of the outermost circle that covers a system level topology and behavior. A key tool developed along the way was a compact model for GNRs that capture chemical, geometric, electronic, electrostatic and transport properties through a set of simple parameterized equations calibrated against a few emerging experiments. Such a Y-chart is the first step towards a holistic approach that we hope will catapult graphene from the domain of fascinating physics and chemistry to technologically relevant electronic applications.

\section*{18.6. Acknowledgements}

We would like to thank Keith Williams, Kurt Gaskill, Jeong-Sun Moon and 
Mark Lundstrom for useful discussions. This work was supported by
the NSF-NRI, NSF-NIRT and the UVa-FEST grants. 






\bibliographystyle{spphys.bst}
\bibliography{graphene_review}
%


\printindex
\end{document}